\documentclass[a4paper,12pt]{article}
\usepackage[T2A]{fontenc} 
\usepackage[russian, english]{babel} 
\usepackage[dvips]{graphicx} 
\makeindex 
\usepackage{epsfig} 
\usepackage[dvips]{texdraw} \input{txdtools}
\usepackage{picinpar} 
\usepackage{amsmath}
\usepackage{amssymb}
\usepackage{indentfirst} 
\allowdisplaybreaks

\begin{document}
\begin{center}

{\bf \large $K^0_{l3\gamma}$ decays:\\
\vspace{2mm} branching ratios and $T$-odd momenta correlations}

\vspace{3mm}

A.S.\,Rudenko\footnote{a.s.rudenko@inp.nsk.su}

Budker Institute of Nuclear Physics\\and Novosibirsk State
University,\\ 630090, Novosibirsk, Russia

\end{center}

\begin{abstract}

The branching ratios of the $K^0 \to \pi^- l^+ \nu_l\gamma \hspace{2mm} (l = e,
\mu)$ decays, and the $T$-odd triple momenta correlations
$\xi=\vec{q}\cdot[\vec{p}_l \times \vec{p}_\pi]/M^3_K$, due to the electromagnetic
final state interaction, in these processes are calculated. The contributions on the
order of $\omega^{-1}$ and $\omega^0$ to the corresponding amplitudes are treated
exactly. For the branching ratios and $T$-odd correlation in $K^0 \to \pi^- e^+
\nu_e\gamma$ decay, the corrections on the order of $\omega$ are estimated and
demonstrated to be small. The results for the branching ratios are in good agreement
with the previous ones. The $T$-odd triple momenta correlations in $K^0_{l3\gamma}$
decays are calculated for the first time. The values of the $\xi$-odd asymmetry
constitute $-1\times10^{-4}$ and $-4.5\times10^{-4}$ in the $K^0 \to \pi^- \mu^+
\nu_\mu \gamma$ and $K^0 \to \pi^- e^+ \nu_e\gamma$ decays, respectively.
\end{abstract}

\vspace{5mm}

{\bf 1.} The $K^0 \to \pi^- l^+ \nu_l\gamma \hspace{2mm} (l = e, \mu)$ decays were
previously studied theoretically in \linebreak Refs.~\cite{ffs,ffs1,gas,gkpv}.
Therein the branching ratios of these decays were calculated. As for the $T$-odd
triple momenta correlations $\xi=\vec{q}\cdot[\vec{p}_l \times \vec{p}_\pi]/M^3_K$,
as induced by the electromagnetic final state interaction, they were considered only
in the $K^+ \to \pi^0 l^+ \nu_l\gamma$ decays \cite{brag,mkm,khr}; here and below
$M_K$ is the kaon mass and $\vec{q}$, $\vec{p}_l$, $\vec{p}_\pi$ are the momenta of
$\gamma$, $l$, $\pi$, respectively. In principle, these triple correlations can be
used to probe new $CP$-odd effects beyond the Standard Model, which could also
contribute to them.

In the theoretical analysis of radiative effects in the discussed processes, the
treatment of the accompanying radiation, which gives the effects on the order of
$\omega^{-1}$ and $\omega^0$ (the last ones originate from the radiation due to the
lepton magnetic moment), is straightforward (here and below $\omega$ is the photon
energy). As to the structure radiation contribution on the order of $\omega^0$, it
is also under control, in fact due to the gauge invariance \cite{low}. The
contributions on the order of $\omega$ (and higher) depend directly on the photon
field strength $F_{\mu\nu}$ (and its derivatives) and cannot be fixed in a
model-independent way. We assume that the corrections on the order of $\omega$ and
higher are relatively small. Indeed, more quantitative arguments presented below
demonstrate that such contributions into the discussed branching ratios and into the
$T$-odd momenta correlation in $K^0_{e3\gamma}$ decay do not exceed $20\%$.

\vspace{3mm}

\begin{figure}[h]
\center
\begin{tabular}{c c c c c}
\includegraphics[scale=1.2]{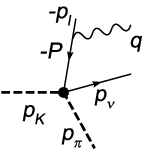} &  &
\includegraphics[scale=1.2]{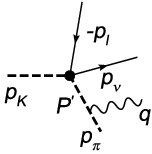} &  &
\includegraphics[scale=1.2]{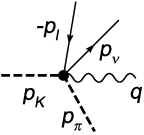} \\
a & & b & & c
\end{tabular}
\caption {The tree diagrams of $K^0 \to \pi^- l^+ \nu_l\gamma $ decays}
\label{fig:1}
\end{figure}

{\bf 2.} At the tree level, the $K^0 \to \pi^- l^+ \nu_l \gamma$ decays are
described by the Feynman graphs in Fig.~\ref{fig:1}.

The matrix elements for diagrams 1a and 1b look as follows:
\begin{multline}\label{M1a}
M_{1a}=\frac{G}{\sqrt{2}}\sin{\theta_c}e[f_+(t)(p_K+p_\pi)_\alpha+f_-(t)(p_K-p_\pi)_\alpha]
\bar{u}_\nu\gamma^\alpha(1+\gamma_5)
\left(\frac{p_le^*}{p_lq}+\frac{\hat{q}\hat{e}^*}{2p_lq}\right)v_l,
\end{multline}
\begin{equation}\label{M1b}
M_{1b}=-\frac{G}{\sqrt{2}}\sin{\theta_c}e[f_+(t')(p_K+p_\pi+q)_\alpha+f_-(t')(p_K-p_\pi-q)_\alpha]
\bar{u}_\nu\gamma^\alpha(1+\gamma_5)v_l\frac{p_\pi
e^*}{p_\pi q};
\end{equation}

\noindent here $G$ is the Fermi coupling constant, $\theta_c$ is the Cabibbo angle,
$e$ is the elementary charge ($e>0$), $t=(p_K-p_\pi)^2$, $t'=(p_K-p_\pi-q)^2$; here
and below the lower indices attached to the matrix elements match the corresponding
Feynman diagrams in the figures.

Usually the dependence of the form factors $f_+$ and $f_-$ on the momentum transfer $t$ is described by
formula
\begin{equation} \label{f-}
f_\pm(t)=f_\pm(0)\left(1+\lambda_\pm\frac{t}{m^2_\pi}\right).
\end{equation}
The experimental data are adequately described by Eq.~(\ref{f-}) with
$\lambda_+\approx 0.03$ for $l=\mu$ and $l=e$; $\lambda_-=0$ for $l=\mu$
\cite{PDG10}; $\lambda_-$ for $l=e$ is unknown, but one may assume that it is also
close to zero.

In the $K^0_{l3\gamma}$ decays the ratio $\lambda_\pm t/m^2_\pi$ is small,
$\lambda_\pm t/m^2_\pi \lesssim 0.1$, so one can put $f_\pm(t)=f_\pm(0)$. Since the
ratio $\xi(0)=f_-(0)/f_+(0)\lesssim 0.1$ is also small \cite{PDG04}, one can neglect
$f_-(0)$ with the same accuracy.

Thus, our expressions (\ref{M1a}) and (\ref{M1b}) simplify to

\begin{equation}
M_{1a}=\frac{G}{\sqrt{2}}\sin{\theta_c}ef_+(0)(p_K+p_\pi)_\alpha\bar{u}_\nu\gamma^\alpha(1+\gamma_5)
\left(\frac{p_le^*}{p_lq}+\frac{\hat{q}\hat{e}^*}{2p_lq}\right)v_l,
\end{equation}

\begin{equation}
M_{1b}=-\frac{G}{\sqrt{2}}\sin{\theta_c}ef_+(0)(p_K+p_\pi+q)_\alpha
\bar{u}_\nu\gamma^\alpha(1+\gamma_5)v_l\frac{p_\pi e^*}{p_\pi q}.
\end{equation}

However, the sum of diagrams 1a and 1b is not gauge invariant: it does not vanish
under the substitution $e^*\to q$. To restore the gauge invariance, one should add
the third diagram where a photon is directly emitted from the vertex (see
Fig.~\ref{fig:1}c). This contact amplitude has no single-particle intermediate
states and therefore is on the order of $\omega^0$ and higher. The contribution
$\sim\omega^0$, as derived with the Low technique \cite{low}, is

\begin{equation}
M_{1c}=\frac{G}{\sqrt{2}}\sin{\theta_c}ef_+(0)
e^*_\alpha\bar{u}_\nu\gamma^\alpha(1+\gamma_5)v_l.
\end{equation}

Thus, the model-independent gauge invariant tree amplitude of $K^0_{l3\gamma}$
decays, including only terms on the order of $\omega^{-1}$ and $\omega^0$ (but all
of them!), is
\begin{multline} \label{m1}
M_{tree}=M_{1a}+M_{1b}+M_{1c}=\frac{G}{\sqrt{2}}\sin{\theta_c}ef_+(0)\left\{(p_K+p_\pi)_\alpha
\bar{u}_\nu\gamma^\alpha(1+\gamma_5)v_l
\left(\frac{p_le^*}{p_lq}-\frac{p_\pi e^*}{p_\pi q}\right) \right. \\
\left.
+(p_K+p_\pi)_\alpha\bar{u}_\nu\gamma^\alpha(1+\gamma_5)\frac{\hat{q}\hat{e}^*}{2p_lq}v_l+
\left(e^*_\alpha-\frac{p_\pi e^*}{p_\pi q}q_\alpha\right)
\bar{u}_\nu\gamma^\alpha(1+\gamma_5)v_l\right\}.
\end{multline}
This expression agrees with the corresponding formulas in
Ref.~\cite{gas} (if our $f_+(0)$ is set to its $SU(3)$ value
$f_+(0)=1$\,).

It is convenient to present amplitude (\ref{m1}) as a sum of gauge-invariant
contributions. They are the \textquotedblleft infrared\textquotedblright \ term
$M_{IR}$ corresponding to the sum of the amplitudes of accompanying radiation by the
pion and lepton (independent of the lepton magnetic moment), the magnetic term
$M_{mag}$ which is the amplitude of spin-dependent accompanying radiation of the
lepton magnetic moment, and the Low term $M_{Low}$:

\begin{equation}
M_{IR}=\frac{G}{\sqrt{2}}\sin{\theta_c}ef_+(0)(p_K+p_\pi)_\alpha\bar{u}_\nu\gamma^\alpha(1+\gamma_5)v_l
\left(\frac{p_le^*}{p_lq}-\frac{p_\pi e^*}{p_\pi q}\right),
\end{equation}

\begin{equation}
M_{mag}=\frac{G}{\sqrt{2}}\sin{\theta_c}ef_+(0)(p_K+p_\pi)_\alpha\bar{u}_\nu\gamma^\alpha(1+\gamma_5)\frac{\hat{q}\hat{e}^*}{2p_lq}v_l,
\end{equation}

\begin{equation}
M_{Low}=\frac{G}{\sqrt{2}}\sin{\theta_c}ef_+(0) \left(e^*_\alpha-\frac{p_\pi
e^*}{p_\pi q}q_\alpha\right) \bar{u}_\nu\gamma^\alpha(1+\gamma_5)v_l.
\end{equation}

In the lowest order in $G$ the decay width $\Gamma(K^0 \to \pi^- l^+ \nu_l \gamma)$
is equal to the sum $\Gamma(K^0_L \to \pi^- l^+ \nu_l \gamma)+\Gamma(K^0_L \to \pi^+
l^- \bar{\nu}_l \gamma)$, which is measured experimentally (see also \cite{gkpv}).
The results of calculation and experimental values for $K^0_L \to \pi^\pm l^\mp
\nu_l \gamma$ branching ratios are presented in Table~\ref{table:1}; here the
following cuts in the kaon rest frame are used: $\omega \geqslant 30$ MeV and
$\theta_{l\gamma}\geqslant 20^\circ$ (except the experimental results for $l=\mu$,
where only restriction $\omega \geqslant 30$ MeV is imposed).

We use in calculation the particle masses $m_K=497.6$ MeV, $m_\pi=139.57$ MeV,
$m_\mu=105.658$ MeV, and $m_e=0.511$ MeV, the Fermi coupling constant
$G=1.1664\times 10^{-11}$ MeV$^{-2}$, the fine-structure constant
$e^2/4\pi=\alpha=1/137$, the Planck constant $\hbar=6.582\times 10^{-22}$
MeV$\cdot$s, $K^0_L$ mean life $\tau=5.1\times 10^{-8}$ s,
$\sin{\theta_c}f_+(0)=0.217$.

\begin{table*}[h]
\begin{center}
\renewcommand{\arraystretch}{1.3}
\begin{tabular}{|c|c|c|} \hline
 & $l=\mu$ & $l=e$ \\
\hline
Bijnens \em{et al}. \rm{\cite{gas}} & $5.2\times 10^{-4}$ & $3.6\times 10^{-3}$\\
\hline
present work & $5.00\times 10^{-4}$ & $3.45\times 10^{-3}$ \\
\hline experimental values \cite{PDG10} & $(5.65\pm0.23)\times 10^{-4}$ & $(3.79\pm0.06)\times 10^{-3}$\\
& $\omega \geqslant 30$ MeV &\\\hline
\end{tabular}
\caption{Branching ratios of $K^0_L \to \pi^\pm l^\mp \nu_l \gamma$ decays}
\label{table:1}
\end{center}
\end{table*}

The accuracy of our results can be estimated as follows. The leading corrections to
them are due to the structure radiation from the hadronic vertex. They are
proportional to the photon field strength, i.e.\ are on the order of $\omega$. There
are good reasons to believe that these corrections are less than the Low structure
amplitudes which are on the order of $\omega^0$. The Low contributions (including of
course their interference with the accompanying radiation) to the discussed $K^0_L
\to \pi^\pm \mu^\mp \nu_\mu \gamma$ and $K^0_L \to \pi^\pm e^\mp \nu_e \gamma$
branching ratios, according to our calculations, constitute $0.79 \times 10^{-4}$
and $0.27\times10^{-3}$, respectively. Let us note also that corrections to the
quoted results derived in Ref.~\cite{gas} in the chiral perturbation theory are of
similar magnitude.

As mentioned, additional corrections on the level of 20\% to the branching ratios
originate from our neglect of the form factor $f_-(t)$ and of the $t$-dependence of
$f_+(t)$ (see also Table I in Ref.~\cite{ffs1}). As to the relative accuracy of our
numerical integration over phase space of final particles, it is about $10^{-3}$.

To compare properly our results with those of Ref.~\cite{gas} one should keep in
mind that now the experimental values of some quantities are known with better
accuracy. Indeed, we use $\sin{\theta_c}f_+(0)=0.217$ in our calculation and, as far
as we can see, in Ref.~\cite{gas} the corresponding value is $0.22$. Substitution of
one of these values for another alters the results by about $3\%$.

Thus, our results for the branching ratios agree reasonably well
with those of Ref.~\cite{gas}.

\vspace{1cm}

{\bf 3.} The $T$-odd triple momenta correlations $\xi=\vec{q}\cdot[\vec{p}_l \times
\vec{p}_\pi]/M^3_K$ in the $K^0 \to \pi^- l^+ \nu_l \gamma$ decays arise from the
interference term $2\mathrm{Re}(M^\dagger_{tree} A_{loop})$ in the decay rate; here
$M_{tree}$ is the tree amplitude and $A_{loop}$ is the anti-Hermitian part of the
loop diagrams presented below. As we are interested in the effect due to the
electromagnetic final state interaction we consider only the one-loop diagrams
generated from the tree ones by attaching the virtual photon. The on-mass-shell
intermediate particles on the diagrams are marked by crosses.

The anti-Hermitian part of the sum of one-loop diagrams is written as
\begin{equation}
A_{loop}=\frac{i}{8\pi^2}\sum_n M_{fn}M^*_{in},
\end{equation}
where the sum over $n$ includes the summation over the polarizations and the
integration over the phase space of intermediate particles.

It is natural to divide the loop diagrams into four groups
according to the type of amplitudes $M_{fn}$ and $M^*_{in}$.

\begin{figure}[h]
\center
\begin{tabular}{c c c c c}
\includegraphics[scale=1.2]{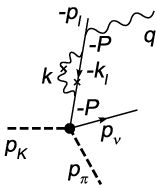} & &
\includegraphics[scale=1.2]{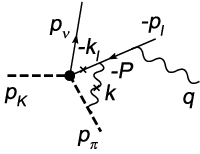} & &
\includegraphics[scale=1.2]{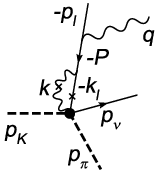} \\
\includegraphics[scale=1.2]{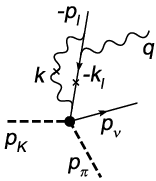} & &
\includegraphics[scale=1.2]{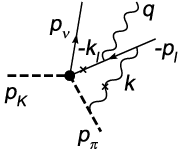} & &
\includegraphics[scale=1.2]{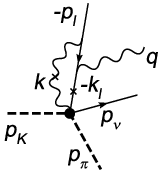} \\
\end{tabular}
\caption {The one-loop diagrams (group I -- the Compton scattering off the lepton)}
\label{fig:2}
\end{figure}

\begin{figure}[h]
\center
\begin{tabular}{c c}
\includegraphics[scale=1.2]{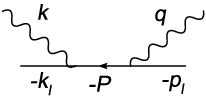} &
\includegraphics[scale=1.2]{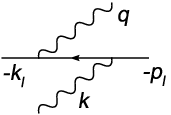} \\
a & b
\end{tabular}
\caption {The Compton scattering off the lepton} \label{fig:3}
\end{figure}

In the first group (see Fig.~\ref{fig:2}) the amplitude $M_{fn}$
depicts the Compton scattering off the intermediate lepton (see
Fig.~\ref{fig:3}):

\begin{equation}
M_{fn}=M_{3a}+M_{3b}=e^2\bar{v}_{k_l} \hat{e}_k \frac{\hat{P}-m_l}{2p_{l}q}
\hat{e}^* v_l + e^2\bar{v}_{k_l} \hat{e}^* \frac{\hat{p}_l-\hat{k}-m_l}{-2p_{l}k}
\hat{e}_k v_l.
\end{equation}

As to $M^*_{in}$, it is the same tree amplitude (\ref{m1}), up to the change of some
notations:
\begin{multline}
M^*_{in}=M_{ni}=M_{tree}\rvert_{\substack{q\to k,\\ p_l\to k_l,\\ e\to e_k}}=
\frac{G}{\sqrt{2}}\sin{\theta_c}ef_+(0)\left\{(p_K+p_\pi)_\alpha\bar{u}_\nu\gamma^\alpha(1+\gamma_5)v_{k_l}
\left(\frac{k_l e^*_k}{p_l q}-\frac{p_\pi e^*_k}{p_\pi k}\right) \right. \\
\left.
+(p_K+p_\pi)_\alpha\bar{u}_\nu\gamma^\alpha(1+\gamma_5)\frac{\hat{k}\hat{e}^*_k}{2p_l
q}v_{k_l}+ \left(e^*_{k\alpha}-\frac{p_\pi e^*_k}{p_\pi k}k_\alpha\right)
\bar{u}_\nu\gamma^\alpha(1+\gamma_5)v_{k_l}\right\}.
\end{multline}

The element of phase space is
\begin{equation}
d\rho=\frac{d^3k}{2\omega_k}\frac{d^3k_l}{2\omega_l}\delta^{(4)}(k_l+k-P),
\end{equation}
where $P=p_l+q$.

\begin{figure}[h]
\center
\begin{tabular}{c c c}
\includegraphics[scale=1.2]{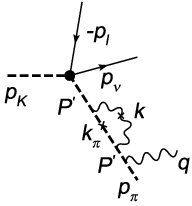} &
\includegraphics[scale=1.2]{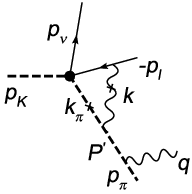} &
\includegraphics[scale=1.2]{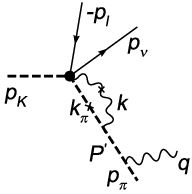} \\
\includegraphics[scale=1.2]{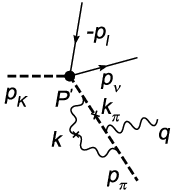} &
\includegraphics[scale=1.2]{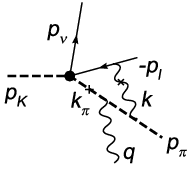} &
\includegraphics[scale=1.2]{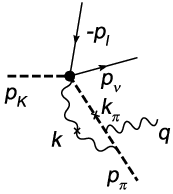} \\
\includegraphics[scale=1.2]{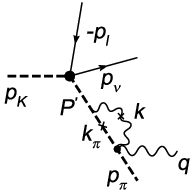} &
\includegraphics[scale=1.2]{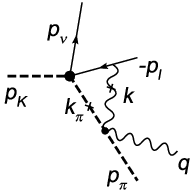} &
\includegraphics[scale=1.2]{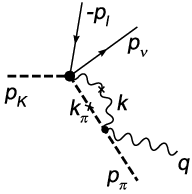}
\end{tabular}
\caption {The one-loop diagrams (group II -- the Compton scattering off the
$\pi$-meson)} \label{fig:4}
\end{figure}

\begin{figure}[h]
\center
\begin{tabular}{c c c}
\includegraphics[scale=1.2]{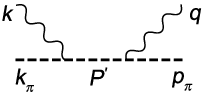} &
\includegraphics[scale=1.2]{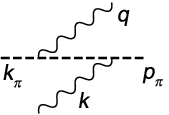} &
\includegraphics[scale=1.2]{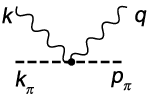}\\
a & b & c
\end{tabular}
\caption {The Compton scattering off the $\pi$-meson}
\label{fig:5}
\end{figure}

In the second group of one-loop diagrams (see Fig.~\ref{fig:4})
the amplitude $M_{fn}$ corresponds to the Compton scattering off
the $\pi$-meson (see Fig.~\ref{fig:5}):

\begin{equation}
M_{fn}=M_{5a}+M_{5b}+M_{5c}=2e^2\left\{-\frac{(p_\pi e^*)(k_\pi e_k)}{p_\pi
q}+\frac{(p_\pi e_k)(k_\pi e^*)}{p_\pi k}+(e_k e^*)\right\}.
\end{equation}

The amplitude $M^*_{in}$ is the tree amplitude (\ref{m1}) up to the change of
notations:
\begin{multline}
M^*_{in}=M_{ni}=M_{tree}\rvert_{\substack{q\to k,\\ p_{\pi}\to k_{\pi},\\ e\to
e_k}}=
\frac{G}{\sqrt{2}}\sin{\theta_c}ef_+(0)\left\{(p_K+k_\pi)_\alpha\bar{u}_\nu\gamma^\alpha(1+\gamma_5)v_l
\left(\frac{p_l e^*_k}{p_l k}-\frac{k_\pi e^*_k}{p_\pi q}\right) \right. \\
\left.
+(p_K+k_\pi)_\alpha\bar{u}_\nu\gamma^\alpha(1+\gamma_5)\frac{\hat{k}\hat{e}^*_k}{2p_l
k}v_l+ \left(e^*_{k\alpha}-\frac{k_\pi e^*_k}{p_\pi q}k_\alpha\right)
\bar{u}_\nu\gamma^\alpha(1+\gamma_5)v_l\right\}.
\end{multline}

The element of phase space is
\begin{equation}
d\rho'=\frac{d^3k}{2\omega_k}\frac{d^3k_\pi}{2\omega_\pi}\delta^{(4)}(k_\pi+k-P'),
\end{equation}
where $P'=p_\pi+q$.

\begin{figure}[h]
\center
\begin{tabular}{c c c c c}
& \includegraphics[scale=1.2]{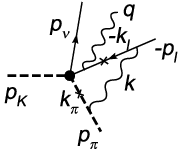} &
\includegraphics[scale=1.2]{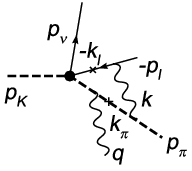} &
\includegraphics[scale=1.2]{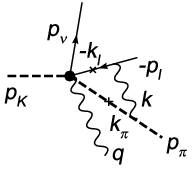} &
\end{tabular}
\caption {The one-loop diagrams (group III -- $\pi$-$l$ scattering)} \label{fig:6}
\end{figure}

\begin{figure}[h]
\center
\begin{tabular}{c}
\includegraphics[scale=1.2]{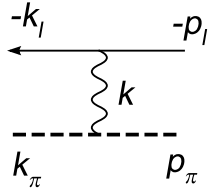}
\end{tabular}
\caption {The $\pi$-$l$ scattering diagram} \label{fig:7}
\end{figure}

In the third group of one-loop diagrams (see Fig.~\ref{fig:6}) the amplitude
$M_{fn}$ is $\pi$-$l$ scattering amplitude (see Fig.~\ref{fig:7}):

\begin{equation}
M_{fn}=M_7=e^2|F_\pi(k^2)|\frac{1}{k^2}(p_\pi+k_\pi)_\alpha
\bar{v}_{k_l}\gamma^\alpha v_l.
\end{equation}

The amplitude $M^*_{in}$ is also the tree amplitude (\ref{m1}) up to the change of
notations:
\begin{multline}
M^*_{in}=M_{ni}=M_{tree}\rvert_{\substack{p_l \to k_l,\\ p_\pi \to k_\pi}}=
\frac{G}{\sqrt{2}}\sin{\theta_c}ef_+(0)\left\{(p_K+k_\pi)_\alpha\bar{u}_\nu\gamma^\alpha(1+\gamma_5)v_{k_l}
\left(\frac{k_l e^*}{k_lq}-\frac{k_\pi e^*}{k_\pi q}\right) \right. \\
\left.
+(p_K+k_\pi)_\alpha\bar{u}_\nu\gamma^\alpha(1+\gamma_5)\frac{\hat{q}\hat{e}^*}{2k_l
q}v_{k_l}+ \left(e^*_{\alpha}-\frac{k_\pi e^*}{k_\pi q}q_\alpha\right)
\bar{u}_\nu\gamma^\alpha(1+\gamma_5)v_{k_l}\right\}.
\end{multline}

The element of phase space is
\begin{equation}
d\rho''=\frac{d^3k_l}{2\omega_l}\frac{d^3k_\pi}{2\omega_\pi}\delta^{(4)}(k_l+k_\pi-P''),
\end{equation}
where $P''=p_l+p_\pi$.

\begin{figure}[h]
\center
\begin{tabular}{c}
\begin{tabular}{c c}
\includegraphics[scale=1.2]{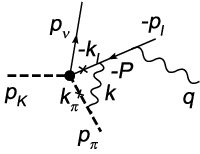} &
\includegraphics[scale=1.2]{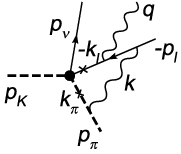}
\end{tabular} \\
\begin{tabular}{c c c}
\includegraphics[scale=1.2]{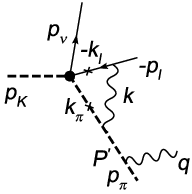} &
\includegraphics[scale=1.2]{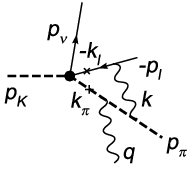} &
\includegraphics[scale=1.2]{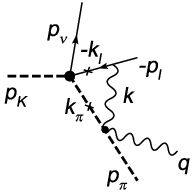}
\end{tabular}
\end{tabular}
\caption {The one-loop diagrams (group IV -- $\pi$-$l$ scattering with emission of a
photon)} \label{fig:8}
\end{figure}

\begin{figure}[h]
\center
\begin{tabular}{c}
\begin{tabular}{c c}
\includegraphics[scale=1.2]{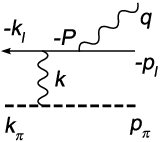} &
\includegraphics[scale=1.2]{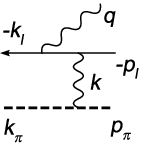}\\
a & b
\end{tabular} \\
\begin{tabular}{c c c}
\includegraphics[scale=1.2]{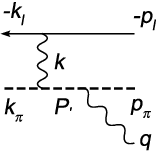} &
\includegraphics[scale=1.2]{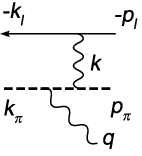} &
\includegraphics[scale=1.2]{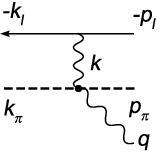}\\
c & d & e
\end{tabular}
\end{tabular}
\caption {$\pi$-$l$ scattering with emission of a photon} \label{fig:9}
\end{figure}

And finally, in the fourth group of one-loop diagrams (see Fig.~\ref{fig:8}) the
amplitude $M_{fn}$ is $\pi$-$l$ scattering with emission of a photon (see
Fig.~\ref{fig:9}):

\begin{multline}
M_{fn}=M_{9a}+M_{9b}+M_{9c}+M_{9d}+M_{9e}=e^3|F_\pi(k^2)|\frac{1}{k^2}\left\{(p_\pi+k_\pi)_\alpha
\bar{v}_{k_l}\gamma^\alpha\left(\frac{p_l e^*}{p_l q}+\frac{\hat{q}\hat{e}^*}{2p_l
q}\right) v_l \right.\\
-(p_\pi+k_\pi)_\alpha\bar{v}_{k_l}\left(\frac{k_l e^*}{k_l
q}+\frac{\hat{q}\hat{e}^*}{2k_l q}\right)\gamma^\alpha v_l-(p_\pi+k_\pi+q)_\alpha
\bar{v}_{k_l} \gamma^\alpha v_l
\frac{p_\pi e^*}{p_\pi q}\\
\left. +(p_\pi+k_\pi-q)_\alpha \bar{v}_{k_l} \gamma^\alpha v_l \frac{k_\pi
e^*}{k_\pi q}+2e^*_\alpha\bar{v}_{k_l} \gamma^\alpha v_l \right\}.
\end{multline}

\begin{figure}[h]
\center
\begin{tabular}{c}
\includegraphics[scale=1.2]{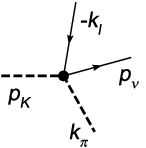}
\end{tabular}
\caption {$K^0_{l3}$ decays} \label{fig:10}
\end{figure}

As to $M^*_{in}$, it is the amplitude of $K^0_{l3}$ decays (see Fig.~\ref{fig:10}):

\begin{equation}
M^*_{in}=M_{ni}=M_{10}=\frac{G}{\sqrt{2}}\sin{\theta_c}ef_+(0)
(p_K+k_\pi)_\alpha\bar{u}_\nu\gamma^\alpha(1+\gamma_5)v_{k_l}.
\end{equation}

The element of phase space is
\begin{equation}
d\rho'''=\frac{d^3k_l}{2\omega_l}\frac{d^3k_\pi}{2\omega_\pi}\delta^{(4)}(k_l+k_\pi-P'''),
\end{equation}
where $P'''=p_l+p_\pi+q$.

In the discussed decays $k^2<0$ and
$\sqrt{-k^2}<m_K(1-m^2_\pi/m^2_K)\lesssim 450$ MeV. At these
energies the pion electromagnetic form factor can be approximated
as
\begin{equation}
|F_\pi(k^2)|\approx 1+<r^2_\pi>k^2/6,
\end{equation}
here $<r^2_\pi>$ is the pion mean square charge radius. According to our
calculations (we use the PDG value $<r^2_\pi>=(0.67\,\mathrm{fm})^2$ \cite{PDG10}),
the contribution of term $<r^2_\pi>k^2/6$ to the momenta correlation is small,
namely, it does not exceed 5\% of the contribution of 1. Therefore, we can put
$|F_\pi(k^2)|=1$. Moreover, keeping $<r^2_\pi>$ term itself is beyond the accuracy
of our calculation, because we have already neglected the form factor $f_-(t)$ and
the $t$-dependence of $f_+(t)$.

The one-loop diagrams from group III contain infrared divergent terms, which can be
omitted because infrared corrections to the decay width are factorized and,
therefore, do not contribute to the effect.

The details of the calculation of the $T$-odd correlation
$\xi=\vec{q}\cdot[\vec{p}_l \times \vec{p}_\pi]/M^3_K$ can be found in Appendices.

In fact, what is really measured experimentally is not the $T$-odd triple momenta
correlation $\xi$ by itself but the asymmetry
\begin{equation}
\begin{split}
A_\xi&=\frac{N_+-N_-}{N_++N_-}=\frac{\int \left(|M|^2_{even}+|M|^2_{odd}\right)
d\Phi_{\xi>0}-\int \left(|M|^2_{even}+|M|^2_{odd}\right) d\Phi_{\xi<0}}{\int
\left(|M|^2_{even}+|M|^2_{odd}\right) d\Phi_{\xi>0}+\int
\left(|M|^2_{even}+|M|^2_{odd}\right) d\Phi_{\xi<0}} \\
&=\frac{\int |M|^2_{odd}d\Phi_{\xi>0}}{\int
|M|^2_{even}d\Phi_{\xi>0}}\approx\frac{\int |M|^2_{odd}d\Phi_{\xi>0}}{\int
|M_{tree}|^2d\Phi_{\xi>0}}
\end{split}
\end{equation}
induced by this correlation; here $N_+$ and $N_-$ are the numbers of events with
$\xi>0$ and $\xi<0$, $|M|^2_{even}$ and $|M|^2_{odd}$ are the $\xi$-even and
$\xi$-odd terms in the amplitude squared, and integration is performed over the
phase space of the final particles.

The results for the asymmetry $A_\xi$ in $K^0 \to \pi^- l^+ \nu_l \gamma$ decays are
presented in Table~\ref{table:2} (here $K^0$ means both $K^0_L$ and $K^0_S$, and the
following cuts in the kaon rest frame are used: $\omega \geqslant 30$ MeV and
$\theta_{l\gamma}\geqslant 20^\circ$).

\begin{table*}[h]
\begin{center}
\renewcommand{\arraystretch}{1.3}
\begin{tabular}{|c|c|c|}
\hline
 & $l=\mu$ & $l=e$ \\
\hline
group I ($l-\gamma$) & $-0.54\times 10^{-4}$ & $-1.32\times 10^{-4}$ \\
\hline group II ($\pi-\gamma$) & $-3.6\times 10^{-4}$ & $-3.2\times 10^{-4}$ \\
\hline group III ($\pi-l$) & $1.73\times 10^{-3}$ & $8.6\times 10^{-4}$ \\
\hline group IV ($\pi-l-\gamma$) & $-1.41\times 10^{-3}$ & $-8.6\times 10^{-4}$ \\
\hline total & $-1\times 10^{-4}$ & $-4.5\times 10^{-4}$ \\
\hline
\end{tabular}
\caption{$A_\xi$ in $K^0 \to \pi^- l^+ \nu_l \gamma$ decays}\label{table:2}

\hspace{15mm} ($\omega \geqslant 30$ MeV and $\theta_{l\gamma}\geqslant 20^\circ$)
\end{center}
\end{table*}

The contributions of the Low term $M_{Low}$ to the asymmetry $A_\xi$ in $K^0 \to
\pi^- \mu^+ \nu_\mu \gamma$ and $K^0 \to \pi^- e^+ \nu_e \gamma$ decays, according
to our calculations, constitute $0.9\times 10^{-4}$ and $-0.9\times 10^{-4}$,
respectively. Therefore, the contribution of $M_{Low}$ for $K^0_{e3\gamma}$ decay is
about 20\%. For $K^0_{\mu 3\gamma}$ decay, however, it is large, comparable to the
contributions of the accompanying radiation which are on the order of $\omega^{-1}$
and $\omega^0$. So, it is difficult to estimate reliably the relative magnitude of
the structure radiation contribution proportional to $\omega$, i.e.\ to estimate
reliably the true accuracy of thus-derived result for $A_\xi$ in $K^0 \to \pi^-
\mu^+ \nu_\mu \gamma$ decay. However, we note here that corrections to the discussed
asymmetry in the $K^+ \to \pi^0 l^+ \nu_l\gamma$ decays, derived in Ref.~\cite{mkm}
within the chiral perturbation theory, are small, on the order of 1\%.

The asymmetry $A_\xi$ was measured only in the $K^+_{l3\gamma}$ decays
\cite{akim,tchik}. The experimental errors are on the level of $10^{-1}$ for $l=\mu$
and $10^{-2}$ for $l=e$. Therefore, at present it is impossible to test the
discussed theoretical predictions that are on the order of $10^{-4}$
\cite{brag,khr}.

In any case experimental study of the $T$-odd asymmetry $A_\xi$ in the
$K_{l3\gamma}$ decays would be very important. It would allow one either to test
$CP$-odd effects beyond the Standard Model or to set limits on them.

\subsection*{Acknowledgements}

I am grateful to I.B.\,Khriplovich for his interest in the work and useful
discussions as well as for critical reading of the text, to L.V.\,Kardapoltsev for
his advice concerning numerical calculations, and to R.N.\,Lee and V.S.\,Fadin for
pointing out the necessity to check the initial results.

The work was supported in part by the Russian Foundation for Basic Research through
Grant No.\,11-02-00792-а, by the Federal Program \textquotedblleft Personnel of
Innovational Russia\textquotedblright \ through Grant No.\,14.740.11.0082, and by
Dmitry Zimin's Dynasty Foundation.

\appendix
\subsection*{Appendix A}

\setcounter{equation}{0}
\renewcommand{\theequation}{A.\arabic{equation}}

In this Appendix the list of the integrals that contribute to $A_{loop}$ is given.

\noindent In formulas below we use the following notations:
\[
I_{l\pi}=\sqrt{(p_lp_\pi)^2-m^2_lm^2_\pi},
\]
\[
I_{\pi P}=\sqrt{(p_\pi P)^2-m^2_\pi P^2},
\]
\[
I'_{lP}=\sqrt{(p_lP')^2-m^2_l{P'}^2},
\]
\[
I_{l\pi q}=\sqrt{(p_lp_\pi +p_lq+p_\pi q)^2-m^2_lm^2_\pi}.
\]

We start with the integrals that contribute to group I of the one-loop diagrams:

\begin{equation}
\int d\rho=a_0=\frac{\pi (p_lq)}{P^2};
\end{equation}

\begin{equation}
\int k_\mu d\rho=a_P P_\mu, \mspace{5mu} \mathrm{where} \mspace{10mu}
a_P=\frac{p_lq}{P^2}a_0;
\end{equation}

\begin{equation}
\int \frac{d\rho}{p_lk}=b_0=
\frac{\pi}{2p_lq}\mathrm{ln}\left(\frac{P^2}{m^2_l}\right);
\end{equation}

\begin{equation}
\int \frac{k_\mu}{p_lk} d\rho=B_{1\mu}=b_lp_{l\mu}+b_P P_\mu\,,
\end{equation}
where $b_l$ and $b_P$ are the solutions of the set of equations
\begin{align*}
b_lm^2_l+b_P(p_lP)&=a_0,\\
b_l(p_lP)+b_P P^2&=b_0(p_lq);
\end{align*}

\begin{equation}
\int \frac{k_\mu k_\nu}{p_lk} d\rho=B_{2\mu \nu}=b_2g_{\mu
\nu}+b_{ll}p_{l\mu}p_{l\nu}+b_{PP}P_\mu P_\nu+b_{lP}(p_{l\mu}P_\nu+P_\mu p_{l\nu}),
\end{equation}
where $b_2$, $b_{ll}$, $b_{PP}$, and $b_{lP}$ are the solutions of the set of
equations
\begin{align*}
4b_2+b_{ll}m^2_l+b_{PP}P^2+2b_{lP}(p_lP) &= 0,\\
b_2+b_{ll}m^2_l+b_{lP}(p_lP) &= 0,\\
b_{PP}(p_lP)+b_{lP}m^2_l &= a_P,\\
b_2+b_{PP}P^2+b_{lP}(p_lP) &= b_P(p_lq);
\end{align*}

\begin{equation}
\int \frac{d\rho}{p_\pi k}=c_0=\frac{\pi}{2I_{\pi
P}}\;\mathrm{ln}\!\left(\frac{p_\pi P+I_{\pi P}}{p_\pi P-I_{\pi P}}\right);
\end{equation}

\begin{equation}
\int \frac{k_\mu}{p_\pi k} d\rho=C_{1\mu}=c_\pi p_{\pi\mu}+c_P P_\mu,
\end{equation}
where $c_\pi$ and $c_P$ are the solutions of the set of equations
\begin{align*}
c_\pi m^2_\pi+c_P(p_\pi P)&=a_0,\\
c_\pi(p_\pi P)+c_P P^2&=c_0(p_lq);
\end{align*}

\begin{equation}
\int \frac{d\rho}{(p_lk)(p_\pi k)}
=d_0=\frac{\pi}{2(p_lq)I_{l\pi}}\;\mathrm{ln}\!\left(\frac{p_lp_\pi
+I_{l\pi}}{p_lp_\pi -I_{l\pi}}\right);
\end{equation}

\begin{equation}
\int \frac{k_\mu}{(p_lk)(p_\pi k)} d\rho=D_{1\mu}=d_lp_{l\mu}+d_\pi
p_{\pi\mu}+d_PP_\mu,
\end{equation}
where $d_l$, $d_\pi$, and $d_P$ are the solutions of the set of equations
\begin{align*}
d_lm^2_l+d_\pi(p_lp_\pi)+d_P(p_lP)&=c_0,\\
d_l(p_lp_\pi)+d_\pi m^2_\pi+d_P(p_\pi P)&=b_0,\\
d_l(p_lP)+d_\pi(p_\pi P)+d_P P^2&=d_0(p_lq);
\end{align*}

\begin{multline}
\int \frac{k_\mu k_\nu}{(p_lk)(p_\pi k)} d\rho=D_{2\mu \nu}=d_2g_{\mu
\nu}+d_{ll}p_{l\mu}p_{l\nu}+d_{\pi\pi}p_{\pi\mu}p_{\pi\nu}+d_{PP}P_\mu
P_\nu\\+d_{l\pi}(p_{l\mu}p_{\pi\nu}+p_{\pi\mu}p_{l\nu})+d_{lP}(p_{l\mu}P_\nu+P_\mu
p_{l\nu})+d_{\pi P}(p_{\pi\mu}P_\nu+P_\mu p_{\pi\nu}),
\end{multline}
where $d_2$, $d_{ll}$, $d_{\pi\pi}$, $d_{PP}$, $d_{l\pi}$, $d_{lP}$, and $d_{\pi P}$
are the solutions of the set of equations
\begin{align*}
4d_2+d_{ll}m^2_l+d_{\pi\pi}m^2_\pi+d_{PP}P^2+2d_{l\pi}(p_lp_\pi)+2d_{lP}(p_lP)+2d_{\pi P}(p_\pi P)&=0,\\
d_2+d_{ll}m^2_l+d_{l\pi}(p_lp_\pi)+d_{lP}(p_lP)&=0,\\
d_{\pi\pi}(p_lp_\pi)+d_{l\pi}m^2_l+d_{\pi P}(p_lP)&=c_\pi,\\
d_{PP}(p_lP)+d_{lP}m^2_l+d_{\pi P}(p_lp_\pi)&=c_P,\\
d_2+d_{\pi\pi}m^2_\pi+d_{l\pi}(p_lp_\pi)+d_{\pi P}(p_\pi P)&=0,\\
d_{PP}(p_\pi P)+d_{lP}(p_lp_\pi)+d_{\pi P}m^2_\pi&=b_P,\\
d_2+d_{PP}P^2+d_{lP}(p_lP)+d_{\pi P}(p_\pi P)&=d_P(p_lq).
\end{align*}

Note that if in these integrals and sets of equations one changes all indices $\pi$
(related to $\pi$-meson) to indices $K$ (related to $K$-meson), then one gets
integrals and sets of equations arising in calculation of $T$-odd triple momenta
correlations in $K^+_{l3\gamma}$ decays \cite{brag,khr}.

\vspace{2mm}

Most of the integrals that contribute to group II of the one-loop diagrams are those
contributing to group I up to the change of some notations: indices $l$ related to
lepton must be changed by indices $\pi$ related to $\pi$-meson and vice versa.

\begin{equation}
\int d\rho'=a'_0=\frac{\pi (p_\pi q)}{{P'}^2};
\end{equation}

\begin{equation}
\int k_\mu d\rho'=a'_P P'_\mu, \mspace{5mu} \mathrm{where} \mspace{10mu}
a'_P=\frac{p_\pi q}{{P'}^2}a'_0;
\end{equation}

\begin{equation}
\int k_\mu k_\nu d\rho'=A'_{2\mu\nu}=a'_2 g_{\mu\nu}+a'_{PP}P'_\mu P'_\nu\,,
\end{equation}
where $a'_2$ and $a'_{PP}$ are the solutions of the set of equations
\begin{align*}
4a'_2+a'_{PP}{P'}^2&=0,\\
a'_2+a'_{PP}{P'}^2&=a'_P(p_\pi q);
\end{align*}

\begin{equation}
\int \frac{d\rho'}{p_\pi k}=b'_0=\frac{\pi}{2p_\pi
q}\mathrm{ln}\left(\frac{{P'}^2}{m^2_\pi}\right);
\end{equation}

\begin{equation}
\int \frac{k_\mu}{p_\pi k} d\rho'=B'_{1\mu}=b'_\pi p_{\pi\mu}+b'_P P'_\mu\,,
\end{equation}
where $b'_\pi$ and $b'_P$ are the solutions of the set of equations
\begin{align*}
b'_\pi m^2_\pi+b'_P(p_\pi P')&=a'_0,\\
b'_\pi(p_\pi P')+b'_P {P'}^2&=b'_0(p_\pi q);
\end{align*}

\begin{equation}
\int \frac{k_\mu k_\nu}{p_\pi k} d\rho'=B'_{2\mu\nu}=b'_2 g_{\mu\nu}
+b'_{\pi\pi}p_{\pi\mu}p_{\pi\nu}+b'_{PP}P'_\mu P'_\nu+b'_{\pi
P}(p_{\pi\mu}P'_\nu+P'_\mu p_{\pi\nu}),
\end{equation}
where $b'$, $b'_{\pi\pi}$, $b'_{PP}$, and $b'_{\pi P}$ are the solutions of the set
of equations
\begin{align*}
4b'_2+b'_{\pi\pi}m^2_\pi+b'_{PP}{P'}^2+2b'_{\pi P}(p_\pi P')&=0,\\
b'_2+b'_{\pi\pi}m^2_\pi+b'_{\pi P}(p_\pi P')&=0,\\
b'_{PP}(p_\pi P')+b'_{\pi P}m^2_\pi&=a'_P,\\
b'_2+b'_{PP}{P'}^2+b'_{\pi P}(p_\pi P')&=b'_P(p_\pi q);
\end{align*}

\begin{equation}
\int \frac{d\rho'}{p_l k}
=c'_0=\frac{\pi}{2I'_{lP}}\;\mathrm{ln}\!\left(\frac{p_lP'+I'_{lP}}{p_lP'-I'_{lP}}\right);
\end{equation}

\begin{equation}
\int \frac{k_\mu}{p_l k} d\rho'=C'_{1\mu}=c'_l p_{l\mu}+c'_P P'_\mu,
\end{equation}
where $c'_l$ and $c'_P$ are the solutions of the set of equations
\begin{align*}
c'_l m^2_l+c'_P(p_l P')&=a'_0,\\
c'_l(p_l P')+c'_P {P'}^2&=c'_0(p_\pi q);
\end{align*}

\begin{equation}
\int \frac{d\rho'}{(p_lk)(p_\pi k)}=d'_0=\frac{\pi}{2(p_\pi
q)I_{l\pi}}\;\mathrm{ln}\!\left(\frac{p_lp_\pi+I_{l\pi}}{p_lp_\pi-I_{l\pi}}\right);
\end{equation}

\begin{equation}
\int \frac{k_\mu}{(p_lk)(p_\pi k)} d\rho'=D'_{1\mu}=d'_l p_{l\mu}+d'_\pi
p_{\pi\mu}+d'_P P'_\mu,
\end{equation}
where $d'_l$, $d'_\pi$, and $d'_P$ are the solutions of the set of equations
\begin{align*}
d'_l m^2_l+d'_\pi(p_lp_\pi)+d'_P(p_l P')&=b'_0,\\
d'_l(p_lp_\pi)+d'_\pi m^2_\pi+d'_P(p_\pi P')&=c'_0,\\
d'_l (p_l P')+d'_\pi(p_\pi P')+d'_P {P'}^2&=d'_0(p_\pi q);
\end{align*}

\begin{multline}
\int \frac{k_\mu k_\nu}{(p_lk)(p_\pi k)} d\rho'=D'_{2\mu \nu}=d'_2
g_{\mu\nu}+d'_{ll}p_{l\mu}p_{l\nu}+d'_{\pi\pi}p_{\pi\mu}p_{\pi\nu}+d'_{PP}P'_\mu
P'_\nu\\+d'_{l\pi}(p_{l\mu}p_{\pi\nu}+p_{\pi\mu}p_{l\nu})+d'_{lP}(p_{l\mu}P'_\nu+P'_\mu
p_{l\nu})+d'_{\pi P}(p_{\pi\mu}P'_\nu+P'_\mu p_{\pi\nu}),
\end{multline}
where $d'_2$, $d'_{ll}$, $d'_{\pi\pi}$, $d'_{PP}$, $d'_{l\pi}$, $d'_{lP}$, and
$d'_{\pi P}$ are the solutions of the set of equations
\begin{align*}
4d'_2+d'_{ll}m^2_l+d'_{\pi\pi}m^2_\pi+d'_{PP}{P'}^2+2d'_{l\pi}(p_lp_\pi)+2d'_{lP}(p_lP')+2d'_{\pi P}(p_\pi P')&=0,\\
d'_2+d'_{ll}m^2_l+d'_{l\pi}(p_l p_\pi)+d'_{lP}(p_lP')&=0,\\
d'_{\pi\pi}(p_l p_\pi)+d'_{l\pi}m^2_l+d'_{\pi P}(p_lP')&=b'_\pi,\\
d'_{PP}(p_lP')+d'_{lP}m^2_l+d'_{\pi P}(p_lp_\pi)&=b'_P,\\
d'_2+d'_{\pi\pi}m^2_\pi+d'_{l\pi}(p_l p_\pi)+d'_{\pi P}(p_\pi P')&=0,\\
d'_{PP}(p_\pi P')+d'_{lP}(p_lp_\pi)+d'_{\pi P}m^2_\pi &=c'_P,\\
d'_2+d'_{PP}{P'}^2+d'_{lP}(p_lP')+d'_{\pi P}(p_\pi P')&=d'_P(p_\pi q).
\end{align*}

\vspace{2mm}

The integrals arising in calculation of group III of the one-loop diagrams:

\begin{equation}
\int d\rho''=a''_0=\frac{\pi I_{l\pi}}{{P''}^2};
\end{equation}

\begin{equation}
\int k_{l\mu} d\rho''=a''_P P''_\mu, \mspace{5mu} \mathrm{where} \mspace{10mu}
a''_P=\frac{(p_lP'')}{{P''}^2}a''_0;
\end{equation}

\begin{equation}
\int \frac{d\rho''}{k_lq}=b''_0=\frac{\pi}{2P''q}\;
\mathrm{ln}\!\left(\frac{p_lP''+I_{l\pi}}{p_lP''-I_{l\pi}}\right);
\end{equation}

\begin{equation}
\int \frac{k_{l\mu}}{k_lq} d\rho''=B''_{1\mu}=b''_q q_\mu+b''_P P''_\mu,
\end{equation}
where $b''_q$ and $b''_P$ are the solutions of the set of equations
\begin{align*}
b''_P(P''q)&=a''_0,\\
b''_q(P''q)+b''_P {P''}^2&=b''_0(p_lP'');
\end{align*}

\begin{equation}
\int \frac{k_{l\mu} k_{l\nu}}{k_lq} d\rho''=B''_{2\mu \nu}=b''_2
g_{\mu\nu}+b''_{qq}q_\mu q_\nu+b''_{PP}P''_\mu P''_\nu+b''_{qP}(q_\mu
P''_\nu+P''_\mu q_\nu),
\end{equation}
where $b''_2$, $b''_{qq}$, $b''_{PP}$, and $b''_{qP}$ are the solutions of the set
of equations
\begin{align*}
4b''_2+b''_{PP}{P''}^2+2b''_{qP}(P''q)&=m^2_l b''_0,\\
b''_2+b''_{qP}(P''q)&=0,\\
b''_{qq}(P''q)+b''_{qP}{P''}^2&=b''_q(p_lP''),\\
b''_2+b''_{PP}{P''}^2+b''_{qP}(P''q)&=b''_P(p_lP'');
\end{align*}

\begin{equation}
\int \frac{d\rho''}{k_\pi q}=c''_0=\frac{\pi}{2P''q}\;
\mathrm{ln}\!\left(\frac{p_\pi P''+I_{l\pi}}{p_\pi P''-I_{l\pi}}\right);
\end{equation}

\begin{equation}
\int \frac{k_{\pi\mu}}{k_\pi q} d\rho''=C''_{1\mu}=c''_q q_\mu+c''_P P''_\mu,
\end{equation}
where $c''_q$ and $c''_P$ are the solutions of the set of equations
\begin{align*}
c''_P(P''q)&=a''_0,\\
c''_q(P''q)+c''_P {P''}^2&=c''_0(p_\pi P'');
\end{align*}

\begin{equation}
\int \frac{k_{\pi\mu} k_{\pi\nu}}{k_\pi q} d\rho''=C''_{2\mu\nu}=c''_2
g_{\mu\nu}+c''_{qq}q_\mu q_\nu+c''_{PP}P''_\mu P''_\nu+c''_{qP}(q_\mu
P''_\nu+P''_\mu q_\nu),
\end{equation}
where $c''_2$, $c''_{qq}$, $c''_{PP}$, and $c''_{qP}$ are the solutions of the set
of equations
\begin{align*}
4c''_2+c''_{PP}{P''}^2+2c''_{qP}(P''q)&=m^2_\pi c''_0,\\
c''_2+c''_{qP}(P''q)&=0,\\
c''_{qq}(P''q)+c''_{qP}{P''}^2&=c''_q(p_\pi P''),\\
c''_2+c''_{PP}{P''}^2+c''_{qP}(P''q)&=c''_P(p_\pi P'');
\end{align*}

\begin{equation}
\int \frac{d\rho''}{k^2}=\frac{\pi}{4I_{l\pi}}\;
\mathrm{ln}\!\left(\frac{\mu^2 {P''}^2}{4I^2_{l\pi}}\right) \to 0,
\end{equation}
here $\mu^2$ is the cut-off parameter for the logarithmically divergent integral
over $k^2$, and the sign $\to$ means that infrared divergent terms are omitted;

\begin{equation}
\int \frac{k_{l\mu}}{k^2} d\rho'' \to
D''_{1\mu}=\frac{\pi}{2I_{l\pi}}\left(\frac{p_\pi P''}{{P''}^2}p_{l\mu}-\frac{p_l
P''}{{P''}^2}p_{\pi\mu}\right);
\end{equation}

\begin{equation}
\int \frac{d\rho''}{k^2(k_lq)}=\frac{\pi}{4(p_lq)I_{l\pi}}\left[
\mathrm{ln}\!\left(\frac{\mu^2
{P''}^2}{4I^2_{l\pi}}\right)+\mathrm{ln}\!\left(\frac{m^2_l(P''q)^2}{{P''}^2(p_lq)^2}\right)\right]
\to
e''_0=\frac{\pi}{4(p_lq)I_{l\pi}}\;\mathrm{ln}\!\left(\frac{m^2_l(P''q)^2}{{P''}^2(p_lq)^2}\right);
\end{equation}

\begin{equation}
\int \frac{k_{l\mu}}{k^2(k_lq)} d\rho'' \to E''_{1\mu}=e''_l p_{l\mu}+e''_\pi
p_{\pi\mu}+e''_q q_\mu,
\end{equation}
where $e''_l$, $e''_\pi$, and $e''_q$ are the solutions of the set of equations
\begin{align*}
e''_l(p_l p_\pi)+e''_\pi m^2_\pi+e''_q(p_\pi q)&=\frac{1}{2}\;b''_0 + (p_l p_\pi)e''_0,\\
e''_l m^2_l+e''_\pi(p_l p_\pi)+e''_q(p_l q)&=-\frac{1}{2}\;b''_0 + m^2_l e''_0,\\
e''_l(p_l q)+e''_\pi(p_\pi q)&=0;
\end{align*}

\begin{multline}
\int \frac{k_{l\mu} k_{l\nu}}{k^2(k_l q)} d\rho'' \to E''_{2\mu\nu}=e''_2
g_{\mu\nu}+e''_{ll}p_{l\mu}p_{l\nu}+e''_{\pi\pi}p_{\pi\mu}p_{\pi\nu}+e''_{qq}q_\mu
q_\nu\\+e''_{l\pi}(p_{l\mu}p_{\pi\nu}+p_{\pi\mu}p_{l\nu})+e''_{lq}(p_{l\mu}q_\nu+q_\mu
p_{l\nu})+e''_{\pi q}(p_{\pi\mu}q_\nu+q_\mu p_{\pi\nu}),
\end{multline}
where $e''_2$, $e''_{ll}$, $e''_{\pi\pi}$, $e''_{qq}$, $e''_{l\pi}$, $e''_{lq}$, and
$e''_{\pi q}$ are the solutions of the set of equations
\begin{align*}
4e''_2+e''_{ll}m^2_l+e''_{\pi\pi}m^2_\pi+2e''_{l\pi}(p_lp_\pi)+2e''_{lq}(p_lq)+2e''_{\pi
q}(p_\pi q)=m^2_l e''_0,\\
e''_2 m^2_l+e''_{ll}m^4_l+e''_{\pi\pi}(p_l p_\pi)^2+e''_{qq}(p_l
q)^2+2e''_{l\pi}m^2_l(p_l p_\pi)+2e''_{lq}m^2_l(p_lq)+2e''_{\pi
q}(p_l p_\pi)(p_l
q)\\=\frac{I^2_{l\pi}(D''_1q)}{(P''q)^2}-\frac{1}{2}\left(\frac{(p_l
q)(p_l P'')}{P''
q}+m^2_l\right)b''_0 + m^4_l e''_0,\\
e''_2 m^2_\pi+e''_{ll}(p_l
p_\pi)^2+e''_{\pi\pi}m^4_\pi+e''_{qq}(p_\pi
q)^2+2e''_{l\pi}m^2_\pi(p_l p_\pi)+2e''_{lq}(p_l p_\pi)
(p_\pi q)+2e''_{\pi q}m^2_\pi(p_\pi q) \\
=\frac{I^2_{l\pi}(D''_1q)}{(P''q)^2}+\frac{1}{2}\left(\frac{(p_\pi q)(p_l P'')}{P''q} + (p_l p_\pi)\right)b''_0 + (p_l p_\pi)^2 e''_0,\\
e''_2(p_l p_\pi)+e''_{ll}m^2_l(p_l p_\pi)+e''_{\pi\pi}m^2_\pi(p_l
p_\pi)+e''_{qq}(p_l q)(p_\pi q)+e''_{l\pi}(m^2_l m^2_\pi+(p_l p_\pi)^2)\\
+e''_{lq}((p_l p_\pi)(p_l q)+m^2_l(p_\pi q))+e''_{\pi
q}((p_l p_\pi)(p_\pi q)+m^2_\pi(p_l q))\\
=-\frac{I^2_{l\pi}(D''_1q)}{(P''q)^2}+\frac{1}{2}\left(m^2_l-\frac{(p_\pi q)(p_l
P'')}{P''q}\right)b''_0 + m^2_l (p_l p_\pi)e''_0,\\
e''_2+e''_{lq}(p_lq)+e''_{\pi q}(p_\pi q)=0,\\
e''_{ll}(p_lq)+e''_{l\pi}(p_\pi
q)=\frac{\pi}{2I_{l\pi}}\frac{p_\pi P''}{{P''}^2},\\
e''_{\pi \pi}(p_\pi
q)+e''_{l\pi}(p_lq)=-\frac{\pi}{2I_{l\pi}}\frac{p_l P''}{{P''}^2};
\end{align*}

\begin{equation}
\int \frac{d\rho''}{k^2(k_\pi q)}=\frac{\pi}{4(p_\pi
q)I_{l\pi}}\left[\mathrm{ln}\!\left(\frac{\mu^2
{P''}^2}{4I^2_{l\pi}}\right)+\mathrm{ln}\!\left(\frac{m^2_\pi(P''q)^2}{{P''}^2(p_\pi
q)^2}\right)\right] \to f''_0=\frac{\pi}{4(p_\pi
q)I_{l\pi}}\;\mathrm{ln}\!\left(\frac{m^2_\pi(P''q)^2}{{P''}^2(p_\pi
q)^2}\right);
\end{equation}

\begin{equation}
\int \frac{k_{\pi\mu}}{k^2 (k_\pi q)} d\rho'' \to F''_{1\mu}=f''_l p_{l\mu}+f''_\pi
p_{\pi\mu}+f''_q q_\mu,
\end{equation}
where $f''_l$, $f''_\pi$, and $f''_q$ are the solutions of the set of equations
\begin{align*}
f''_l m^2_l+f''_\pi(p_l p_\pi)+f''_q(p_l q)&=\frac{1}{2}\;c''_0 + (p_l p_\pi)f''_0,\\
f''_l(p_l p_\pi)+f''_\pi m^2_\pi+f''_q(p_\pi q)&=-\frac{1}{2}\;c''_0 + m^2_\pi f''_0,\\
f''_l(p_l q)+f''_\pi(p_\pi q)&=0;
\end{align*}

\begin{multline}
\int \frac{k_{\pi\mu} k_{\pi\nu}}{k^2(k_\pi q)} d\rho'' \to F''_{2\mu\nu}=f''_2
g_{\mu\nu}+f''_{ll}p_{l\mu}p_{l\nu}+f''_{\pi\pi}p_{\pi\mu}p_{\pi\nu}+f''_{qq}q_\mu
q_\nu\\+f''_{l\pi}(p_{l\mu}p_{\pi\nu}+p_{\pi\mu}p_{l\nu})+f''_{lq}(p_{l\mu}q_\nu+q_\mu
p_{l\nu})+f''_{\pi q}(p_{\pi\mu}q_\nu+q_\mu p_{\pi\nu}),
\end{multline}
where $f''_2$, $f''_{ll}$, $f''_{\pi\pi}$, $f''_{qq}$, $f''_{l\pi}$, $f''_{lq}$, and
$f''_{\pi q}$ are the solutions of the set of equations
\begin{align*}
4f''_2+f''_{ll}m^2_l+f''_{\pi\pi}m^2_\pi+2f''_{l\pi}(p_lp_\pi)+2f''_{lq}(p_lq)+2f''_{\pi q}(p_\pi q)=m^2_\pi f''_0,\\
f''_2 m^2_l+f''_{ll}m^4_l+f''_{\pi\pi}(p_l p_\pi)^2+f''_{qq}(p_l
q)^2+2f''_{l\pi}m^2_l(p_lp_\pi)+
2f''_{lq}m^2_l(p_lq)+2f''_{\pi q}(p_lp_\pi)(p_lq)\\
=-\frac{I^2_{l\pi}(D''_1q)}{(P''q)^2}+\frac{1}{2}\left(\frac{(p_l q)(p_\pi P'')}{P''q}+p_lp_\pi\right)c''_0 + (p_l p_\pi)^2 f''_0,\\
f''_2
m^2_\pi+f''_{ll}(p_lp_\pi)^2+f''_{\pi\pi}m^4_\pi+f''_{qq}(p_\pi
q)^2+2f''_{l\pi}m^2_\pi(p_lp_\pi)+2f''_{lq}(p_lp_\pi)(p_\pi q)+2f''_{\pi q}m^2_\pi(p_\pi q)\\
=-\frac{I^2_{l\pi}(D''_1q)}{(P''q)^2}-\frac{1}{2}\left(\frac{(p_\pi q)(p_\pi P'')}{P''q}+m^2_\pi\right)c''_0 + m^4_\pi f''_0,\\
f''_2(p_lp_\pi)+f''_{ll}m^2_l(p_lp_\pi)+f''_{\pi\pi}m^2_\pi(p_lp_\pi)+f''_{qq}(p_lq)(p_\pi
q)+f''_{l\pi}(m^2_l
m^2_\pi+(p_lp_\pi)^2)\\+f''_{lq}((p_lp_\pi)(p_lq)+m^2_l(p_\pi
q))+f''_{\pi q}((p_lp_\pi)(p_\pi
q)+m^2_\pi(p_lq))\\
=\frac{I^2_{l\pi}(D''_1q)}{(P''q)^2}+\frac{1}{2}\left(m^2_\pi-\frac{(p_l q)(p_\pi
P'')}{P''q}\right)c''_0+ m^2_\pi (p_l p_\pi)f''_0,\\
f''_2+f''_{lq}(p_lq)+f''_{\pi q}(p_\pi q)=0,\\
f''_{ll}(p_lq)+f''_{l\pi}(p_\pi
q)=-\frac{\pi}{2I_{l\pi}}\frac{p_\pi P''}{{P''}^2},\\
f''_{\pi \pi}(p_\pi q)+f''_{l\pi}(p_lq)=\frac{\pi}{2I_{l\pi}}\frac{p_l
P''}{{P''}^2}.
\end{align*}

\vspace{2mm}

The integrals arising in calculation of group IV of the one-loop diagrams:

\begin{equation}
\int d\rho'''=a'''_0=\frac{\pi I_{l\pi q}}{{P'''}^2};
\end{equation}

\begin{equation}
\int k_{l\mu} d\rho'''=a'''_P P'''_\mu, \mspace{5mu} \mathrm{where} \mspace{10mu}
a'''_P=\frac{p_l P'''+p_\pi q}{{P'''}^2}a'''_0;
\end{equation}

\begin{equation}
\int \frac{d\rho'''}{k_lq}=b'''_0=\frac{\pi}{2P''q}\;
\mathrm{ln}\!\left(\frac{p_lP'''+p_\pi q+I_{l\pi q}}{p_lP'''+p_\pi q-I_{l\pi
q}}\right);
\end{equation}

\begin{equation}
\int \frac{k_{l\mu}}{k_lq} d\rho'''=B'''_{1\mu}=b'''_q q_\mu+b'''_P P'''_\mu,
\end{equation}
where $b'''_q$ and $b'''_P$ are the solutions of the set of equations
\begin{align*}
b'''_P(P''q)&=a'''_0,\\
b'''_q(P''q)+b'''_P {P'''}^2&=b'''_0(p_lP'''+p_\pi q);
\end{align*}

\begin{equation}
\int \frac{k_{l\mu} k_{l\nu}}{k_lq} d\rho'''=B'''_{2\mu \nu}=b'''_2
g_{\mu\nu}+b'''_{qq}q_\mu q_\nu+b'''_{PP}P'''_\mu P'''_\nu+b'''_{qP}(q_\mu
P'''_\nu+P'''_\mu q_\nu),
\end{equation}
where $b'''_2$, $b'''_{qq}$, $b'''_{PP}$, and $b'''_{qP}$ are the solutions of the
set of equations
\begin{align*}
4b'''_2+b'''_{PP}{P'''}^2+2b'''_{qP}(P''q)&=m^2_l b'''_0,\\
b'''_2+b'''_{qP}(P''q)&=0,\\
b'''_{qq}(P''q)+b'''_{qP}{P'''}^2&=b'''_q(p_lP'''+p_\pi q),\\
b'''_2+b'''_{PP}{P'''}^2+b'''_{qP}(P''q)&=b'''_P(p_lP'''+p_\pi q);
\end{align*}

\begin{equation}
\int \frac{d\rho'''}{k_\pi q}=c'''_0=\frac{\pi}{2P''q}\;
\mathrm{ln}\!\left(\frac{p_\pi P'''+p_lq+I_{l\pi q}}{p_\pi P'''+p_lq-I_{l\pi
q}}\right);
\end{equation}

\begin{equation}
\int \frac{k_{\pi\mu}}{k_\pi q} d\rho'''=C'''_{1\mu}=c'''_q q_\mu+c'''_P P'''_\mu,
\end{equation}
where $c'''_q$ and $c'''_P$ are the solutions of the set of equations
\begin{align*}
c'''_P(P''q)&=a'''_0,\\
c'''_q(P''q)+c'''_P {P'''}^2&=c'''_0(p_\pi P'''+p_lq);
\end{align*}

\begin{equation}
\int \frac{k_{\pi\mu} k_{\pi\nu}}{k_\pi q} d\rho'''=C'''_{2\mu\nu}=c'''_2
g_{\mu\nu}+c'''_{qq}q_\mu q_\nu+c'''_{PP}P'''_\mu P'''_\nu+c'''_{qP}(q_\mu
P'''_\nu+P'''_\mu q_\nu),
\end{equation}
where $c'''_2$, $c'''_{qq}$, $c'''_{PP}$, and $c'''_{qP}$ are the solutions of the
set of equations
\begin{align*}
4c'''_2+c'''_{PP}{P'''}^2+2c'''_{qP}(P''q)&=m^2_\pi c'''_0,\\
c'''_2+c'''_{qP}(P''q)&=0,\\
c'''_{qq}(P''q)+c'''_{qP}{P'''}^2&=c'''_q(p_\pi P'''+p_lq),\\
c'''_2+c'''_{PP}{P'''}^2+c'''_{qP}(P''q)&=c'''_P(p_\pi P'''+p_lq);
\end{align*}

\begin{equation}
\int \frac{d\rho'''}{(p_\pi-k_\pi)^2}=d'''_0=\frac{\pi}{4I_{\pi P}}\;
\mathrm{ln}\!\left(\frac{(p_\pi P)(p_lp_\pi+p_lq+p_\pi q)-m^2_\pi(p_lP)-I_{l\pi
q}I_{\pi P}}{(p_\pi P)(p_lp_\pi+p_lq+p_\pi q)-m^2_\pi(p_lP)+I_{l\pi q}I_{\pi
P}}\right);
\end{equation}

\begin{equation}
\int \frac{k_{l\mu}}{(p_\pi-k_\pi)^2} d\rho'''=D'''_{1\mu}=d'''_\pi
p_{\pi\mu}+d'''_P P_\mu,
\end{equation}
where $d'''_\pi$ and $d'''_P$ are the solutions of the set of equations
\begin{align*}
d'''_\pi m^2_\pi +d'''_P(p_\pi P)&=\frac{1}{2}a'''_0+(p_\pi P)d'''_0,\\
d'''_\pi (p_\pi P)+d'''_P P^2&=-\frac{1}{2}a'''_0+(p_l P)d'''_0;
\end{align*}

\begin{equation}
\int \frac{d\rho'''}{(p_l-k_l)^2}=e'''_0=\frac{\pi}{4I'_{lP}}\;
\mathrm{ln}\!\left(\frac{(p_lP')(p_lp_\pi+p_lq+p_\pi q)-m^2_l(p_\pi P')-I_{l\pi
q}I'_{lP}}{(p_lP')(p_lp_\pi+p_lq+p_\pi q)-m^2_l(p_\pi P')+I_{l\pi q}I'_{lP}}\right);
\end{equation}

\begin{equation}
\int \frac{k_{\pi\mu}}{(p_l-k_l)^2} d\rho'''=E'''_{1\mu}=e'''_l p_{l\mu}+e'''_P
P'_\mu,
\end{equation}
where $e'''_l$ and $e'''_P$ are the solutions of the set of equations
\begin{align*}
e'''_l m^2_l +e'''_P(p_l P')&=\frac{1}{2}a'''_0+(p_l P')e'''_0,\\
e'''_l (p_l P')+e'''_P {P'}^2&=-\frac{1}{2}a'''_0+(p_\pi P')e'''_0;
\end{align*}

\begin{equation}
\int \frac{d\rho'''}{(p_\pi-k_\pi)^2(k_lq)}=f'''_0=\frac{\pi}{4(p_lq)I_{l\pi}}\;
\mathrm{ln}\!\left(\frac{p_lp_\pi(p_lp_\pi+p_lq+p_\pi
q)-m^2_lm^2_\pi-I_{l\pi}I_{l\pi q}}{p_lp_\pi(p_lp_\pi+p_lq+p_\pi
q)-m^2_lm^2_\pi+I_{l\pi}I_{l\pi q}}\right);
\end{equation}

\begin{equation}
\int \frac{k_{l\mu}}{(p_\pi-k_\pi)^2(k_lq)} d\rho'''=F'''_{1\mu}=f'''_l
p_{l\mu}+f'''_\pi p_{\pi\mu}+f'''_q q_\mu,
\end{equation}
where $f'''_l$, $f'''_\pi$, and $f'''_q$ are the solutions of the set of equations
\begin{align*}
f'''_l m^2_l+f'''_\pi(p_l p_\pi)+f'''_q(p_l q)&=-\frac{1}{2}b'''_0-d'''_0+(p_lP)f'''_0,\\
f'''_l(p_l p_\pi)+f'''_\pi m^2_\pi+f'''_q(p_\pi q)&=\frac{1}{2}b'''_0+(p_\pi P)f'''_0,\\
f'''_l(p_l q)+f'''_\pi(p_\pi q)&=d'''_0;
\end{align*}

\begin{multline}
\int \frac{k_{l\mu} k_{l\nu}}{(p_\pi-k_\pi)^2(k_l q)} d\rho'''=F'''_{2\mu\nu}=f'''_2
g_{\mu\nu}+f'''_{ll}p_{l\mu}p_{l\nu}+f'''_{\pi\pi}p_{\pi\mu}p_{\pi\nu}+
f'''_{qq}q_\mu q_\nu\\
+f'''_{l\pi}(p_{l\mu}p_{\pi\nu}+p_{\pi\mu}p_{l\nu})+f'''_{lq}(p_{l\mu}q_\nu+q_\mu
p_{l\nu})+f'''_{\pi q}(p_{\pi\mu}q_\nu+q_\mu p_{\pi\nu}),
\end{multline}
where $f'''_2$, $f'''_{ll}$, $f'''_{\pi\pi}$, $f'''_{qq}$, $f'''_{l\pi}$,
$f'''_{lq}$, and $f'''_{\pi q}$ are the solutions of the set of equations
\begin{align*}
\begin{split}
4f'''_2+f'''_{ll}m^2_l+f'''_{\pi\pi}m^2_\pi+2f'''_{l\pi}(p_lp_\pi)+2f'''_{lq}(p_lq)+2f'''_{\pi
q}(p_\pi q)=m^2_lf'''_0,\\
f'''_2 m^2_l+f'''_{ll}m^4_l+f'''_{\pi\pi}(p_l p_\pi)^2+f'''_{qq}(p_l
q)^2+2f'''_{l\pi}m^2_l(p_l p_\pi)+2f'''_{lq}m^2_l(p_lq)+2f'''_{\pi q}(p_l p_\pi)(p_l
q)\\=(p_lP)^2f'''_0-\frac{1}{2}\left(p_lP+\frac{p_lq}{P''q}(p_lP'''+p_\pi
q)\right)b'''_0 -(p_lP)d'''_0\\+\frac{\pi}{2P''q}\left(\frac{p_\pi
P'''}{{P'''}^2}-\frac{p_\pi q}{P''q}\right)I_{l\pi q}
+\frac{1}{2}a'''_0-(D'''_1p_l),\\
f'''_2 m^2_\pi+f'''_{ll}(p_l p_\pi)^2+f'''_{\pi\pi}m^4_\pi+f'''_{qq}(p_\pi
q)^2+2f'''_{l\pi}m^2_\pi(p_l p_\pi)+2f'''_{lq}(p_l p_\pi)
(p_\pi q)+2f'''_{\pi q}m^2_\pi(p_\pi q) \\
=(p_\pi P)^2f'''_0+\frac{1}{2}\left(p_\pi P+\frac{p_\pi q}{P''q}(p_lP'''+p_\pi q)\right)b'''_0
+\frac{\pi}{2P''q}\left(\frac{p_\pi P'''}{{P'''}^2}-\frac{p_\pi q}{P''q}\right)I_{l\pi q},\\
f'''_2(p_l p_\pi)+f'''_{ll}m^2_l(p_l p_\pi)+f'''_{\pi\pi}m^2_\pi(p_l
p_\pi)+f'''_{qq}(p_l q)(p_\pi q)+f'''_{l\pi}(m^2_l m^2_\pi+(p_l p_\pi)^2)\\+
f'''_{lq}((p_l p_\pi)(p_l q)+m^2_l(p_\pi q))+f'''_{\pi q}((p_l p_\pi)(p_\pi q)+m^2_\pi(p_l q))\\
=(p_lP)(p_\pi P)f'''_0+\frac{1}{2}\left(p_lP-\frac{p_\pi q}{P''q}(p_lP'''+p_\pi
q)\right)b'''_0-(p_\pi P)d'''_0\\
-\frac{1}{2}a'''_0-\frac{\pi}{2P''q}\left(\frac{p_\pi P'''}{{P'''}^2}-\frac{p_\pi q}{P''q}\right)I_{l\pi q},\\
f'''_2+f'''_{lq}(p_lq)+f'''_{\pi q}(p_\pi q)=d'''_P,\\
f'''_{ll}(p_lq)+f'''_{l\pi}(p_\pi q)=d'''_P,\\
f'''_{\pi \pi}(p_\pi q)+f'''_{l\pi}(p_lq)=d'''_\pi;
\end{split}
\end{align*}

\begin{equation}
\int \frac{d\rho'''}{(p_l-k_l)^2(k_\pi q)}=g'''_0=\frac{\pi}{4(p_\pi q)I_{l\pi}}\;
\mathrm{ln}\!\left(\frac{p_lp_\pi(p_lp_\pi+p_lq+p_\pi
q)-m^2_lm^2_\pi-I_{l\pi}I_{l\pi q}}{p_lp_\pi(p_lp_\pi+p_lq+p_\pi
q)-m^2_lm^2_\pi+I_{l\pi}I_{l\pi q}}\right);
\end{equation}

\begin{equation}
\int \frac{k_{\pi\mu}}{(p_l-k_l)^2(k_\pi q)}
d\rho'''=G'''_{1\mu}=g'''_l p_{l\mu}+g'''_\pi p_{\pi\mu}+g'''_q
q_\mu,
\end{equation}
where $g'''_l$, $g'''_\pi$, and $g'''_q$ are the solutions of the set of equations
\begin{align*}
g'''_l m^2_l+g'''_\pi(p_l p_\pi)+g'''_q(p_l q)&=\frac{1}{2}c'''_0+(p_l P')g'''_0,\\
g'''_l(p_l p_\pi)+g'''_\pi m^2_\pi+g'''_q(p_\pi q)&=-\frac{1}{2}c'''_0-e'''_0+(p_\pi P')g'''_0,\\
g'''_l(p_l q)+g'''_\pi(p_\pi q)&=e'''_0;
\end{align*}

\begin{multline}
\int \frac{k_{\pi\mu} k_{\pi\nu}}{(p_l-k_l)^2(k_\pi q)}
d\rho'''=G'''_{2\mu\nu}=g'''_2
g_{\mu\nu}+g'''_{ll}p_{l\mu}p_{l\nu}+g'''_{\pi\pi}p_{\pi\mu}p_{\pi\nu}+
g'''_{qq}q_\mu q_\nu\\
+g'''_{l\pi}(p_{l\mu}p_{\pi\nu}+p_{\pi\mu}p_{l\nu})+g'''_{lq}(p_{l\mu}q_\nu+q_\mu
p_{l\nu})+g'''_{\pi q}(p_{\pi\mu}q_\nu+q_\mu p_{\pi\nu}),
\end{multline}
where $g'''_2$, $g'''_{ll}$, $g'''_{\pi\pi}$, $g'''_{qq}$, $g'''_{l\pi}$,
$g'''_{lq}$, and $g'''_{\pi q}$ are the solutions of the set of equations
\begin{align*}
\begin{split}
4g'''_2+g'''_{ll}m^2_l+g'''_{\pi\pi}m^2_\pi+2g'''_{l\pi}(p_lp_\pi)+2g'''_{lq}(p_lq)+2g'''_{\pi
q}(p_\pi q)=m^2_\pi g'''_0,\\
g'''_2 m^2_l+g'''_{ll}m^4_l+g'''_{\pi\pi}(p_l p_\pi)^2+g'''_{qq}(p_l
q)^2+2g'''_{l\pi}m^2_l(p_l p_\pi)+2g'''_{lq}m^2_l(p_lq)+2g'''_{\pi q}(p_l p_\pi)(p_l
q)\\=(p_l P')^2g'''_0+\frac{1}{2}\left(p_l P'+\frac{p_l q}{P''q}(p_\pi P'''+p_l
q)\right)c'''_0
+\frac{\pi}{2P''q}\left(\frac{p_l P'''}{{P'''}^2}-\frac{p_l q}{P''q}\right)I_{l\pi q},\\
g'''_2 m^2_\pi+g'''_{ll}(p_l p_\pi)^2+g'''_{\pi\pi}m^4_\pi+g'''_{qq}(p_\pi
q)^2+2g'''_{l\pi}m^2_\pi(p_l p_\pi)+2g'''_{lq}(p_l p_\pi)
(p_\pi q)+2g'''_{\pi q}m^2_\pi(p_\pi q) \\
=(p_\pi P')^2g'''_0-\frac{1}{2}\left(p_\pi P'+\frac{p_\pi q}{P''q}(p_\pi P'''+p_l
q)\right)c'''_0 -(p_\pi P')e'''_0\\+\frac{\pi}{2P''q}\left(\frac{p_l
P'''}{{P'''}^2}-\frac{p_l q}{P''q}\right)I_{l\pi q}
+\frac{1}{2}a'''_0-(E'''_1p_\pi),\\
g'''_2(p_l p_\pi)+g'''_{ll}m^2_l(p_l p_\pi)+g'''_{\pi\pi}m^2_\pi(p_l
p_\pi)+g'''_{qq}(p_l q)(p_\pi q)+g'''_{l\pi}(m^2_l m^2_\pi+(p_l p_\pi)^2)\\
+g'''_{lq}((p_l p_\pi)(p_l q)+m^2_l(p_\pi q))+g'''_{\pi q}((p_l p_\pi)(p_\pi q)+m^2_\pi(p_l q))\\
=(p_\pi P')(p_l P')g'''_0+\frac{1}{2}\left(p_\pi P'-\frac{p_l q}{P''q}(p_\pi
P'''+p_l q)\right)c'''_0-(p_l P')e'''_0\\
-\frac{1}{2}a'''_0-\frac{\pi}{2P''q}\left(\frac{p_l P'''}{{P'''}^2}-\frac{p_l q}{P''q}\right)I_{l\pi q},\\
g'''_2+g'''_{lq}(p_lq)+g'''_{\pi q}(p_\pi q)=e'''_P,\\
g'''_{\pi \pi}(p_\pi q)+g'''_{l\pi}(p_lq)=e'''_P,\\
g'''_{ll}(p_lq)+g'''_{l\pi}(p_\pi q)=e'''_l.
\end{split}
\end{align*}

\subsection*{Appendix B}

\setcounter{equation}{0}
\renewcommand{\theequation}{B.\arabic{equation}}
In this Appendix we give the formulas for the anti-Hermitian part $A_{loop}$ of the
one-loop diagrams. It can be expressed as follows:

\begin{multline}
A_{loop}=\frac{i}{8\pi^2}\sum_n
M_{fn}M^*_{in}=\frac{i}{8\pi^2}\sum_n
\left[(M_{3a}+M_{3b})(M_{IR}+M_{mag}+M_{Low})\rvert_{\substack{q\to k,\\
p_l\to k_l,\\ e\to e_k}}\right.\\
\left.+(M_{5a}+M_{5b}+M_{5c})(M_{IR}+M_{mag}+M_{Low})\rvert_{\substack{q\to k,\\
p_\pi \to k_\pi,\\ e\to e_k}}\right.\\
\left. +M_7(M_{IR}+M_{mag}+M_{Low})\rvert_{\substack{p_l\to k_l,\\
p_\pi \to k_\pi}}+(M_{9a}+M_{9b}+M_{9c}+M_{9d}+M_{9e})M_{10}\right]\\
=(A_{3a-IR}+A_{3a-mag}+A_{3a-Low}+A_{3b-IR}+A_{3b-mag}+A_{3b-Low})\\
+(A_{5a-IR}+A_{5a-mag}+A_{5a-Low}+A_{5b-IR}+A_{5b-mag}+A_{5b-Low}+A_{5c-IR}+A_{5c-mag}+A_{5c-Low})\\
+(A_{7-IR}+A_{7-mag}+A_{7-Low})+(A_{9a-10}+A_{9b-10}+A_{9c-10}+A_{9d-10}+A_{9e-10}),
\end{multline}
where
\begin{multline}
A_{3a-IR}=\frac{i}{8\pi^2}\sum_n
M_{3a}M_{IR}\rvert_{\substack{q\to k,\\
p_l\to k_l,\\ e\to e_k}}
=\frac{ie^2}{8\pi^2}\frac{G}{\sqrt{2}}\sin{\theta_c}ef_+(0)\bar{u}_\nu(\hat{p}_K+\hat{p}_\pi)(1+\gamma_5)\\
\times
\left[c_0(\hat{P}-m_l)\hat{p}_\pi-\hat{C}_1\hat{p}_\pi+\frac{m_l}{p_lq}(a_0(\hat{P}-m_l)-a_P\hat{P})\right]
\frac{\hat{P}-m_l}{2p_lq} \hat{e}^* v_l;
\end{multline}

\begin{multline}
A_{3a-mag}=\frac{i}{8\pi^2}\sum_n
M_{3a}M_{mag}\rvert_{\substack{q\to k,\\
p_l\to k_l,\\ e\to
e_k}}=\frac{ie^2}{8\pi^2}\frac{G}{\sqrt{2}}\sin{\theta_c}ef_+(0)
\bar{u}_\nu(\hat{p}_K+\hat{p}_\pi)(1+\gamma_5)\\
\times
\left[2a_P(P^2+2m_l\hat{P})\right]\frac{\hat{P}-m_l}{(2p_lq)^2}
\hat{e}^*v_l;
\end{multline}

\begin{multline}
A_{3a-Low}=\frac{i}{8\pi^2}\sum_n
M_{3a}M_{Low}\rvert_{\substack{q\to k,\\
p_l\to k_l,\\ e\to
e_k}}=\frac{ie^2}{8\pi^2}\frac{G}{\sqrt{2}}\sin{\theta_c}ef_+(0)\bar{u}_\nu(1-\gamma_5)\\
\times
\left[2a_0(\hat{P}+2m_l)-2a_P\hat{P}+\hat{C}_1(\hat{P}-m_l)\hat{p}_\pi\right]\frac{\hat{P}-m_l}{2p_lq}
\hat{e}^* v_l;
\end{multline}

\begin{multline}
A_{3b-IR}=\frac{i}{8\pi^2}\sum_n
M_{3b}M_{IR}\rvert_{\substack{q\to k,\\
p_l\to k_l,\\ e\to
e_k}}=\frac{ie^2}{8\pi^2}\frac{G}{\sqrt{2}}\sin{\theta_c}ef_+(0)
\bar{u}_\nu(\hat{p}_K+\hat{p}_\pi)(1+\gamma_5)\\
\times
\frac{1}{2}\left[(\hat{P}-m_l)\hat{e}^*\left(\frac{1}{p_lq}\left((\hat{p}_l-m_l)(b_0\hat{P}-\hat{B}_1)-
\hat{B}_1\hat{P}\right)-d_0(\hat{p}_l-m_l)\hat{p}_\pi+
\hat{D}_1\hat{p}_\pi\right)\right.\\
\left.-\frac{1}{p_lq}\left(\hat{B}_1\hat{e}^*(\hat{p}_l-m_l)\hat{P}-
B_{2\mu\nu}\gamma^\mu\hat{e}^*(\hat{p}_l-m_l)\gamma^\nu-B_{2\mu\nu}\gamma^\mu\hat{e}^*\gamma^\nu\hat{P}\right)
\right.\\
\left.+\hat{D}_1\hat{e}^*(\hat{p}_l-m_l)\hat{p}_\pi-D_{2\mu\nu}\gamma^\mu\hat{e}^*\gamma^\nu\hat{p}_\pi\right]v_l;
\end{multline}

\begin{multline}
A_{3b-mag}=\frac{i}{8\pi^2}\sum_n
M_{3b}M_{mag}\rvert_{\substack{q\to k,\\
p_l\to k_l,\\ e\to
e_k}}=\frac{ie^2}{8\pi^2}\frac{G}{\sqrt{2}}\sin{\theta_c}ef_+(0)
\bar{u}_\nu(\hat{p}_K+\hat{p}_\pi)(1+\gamma_5)\\
\times
\frac{1}{2p_lq}\left[B_{2\mu\nu}\gamma^\mu\hat{p}_l\hat{e}^*\gamma^\nu+4m_lB_{2\mu\nu}\gamma^\mu
e^{*\nu}-\hat{B}_1\hat{p}_l\hat{e}^*\hat{P}-4m_l(p_le^*)\hat{B}_1-m^2_l\hat{B}_1\hat{e}^*\right]v_l;
\end{multline}

\begin{multline}
A_{3b-Low}=\frac{i}{8\pi^2}\sum_n
M_{3b}M_{Low}\rvert_{\substack{q\to k,\\
p_l\to k_l,\\ e\to
e_k}}=\frac{ie^2}{8\pi^2}\frac{G}{\sqrt{2}}\sin{\theta_c}ef_+(0)\bar{u}_\nu(1-\gamma_5)\\
\times
\frac{1}{2}\left[D_{2\mu\nu}\gamma^\mu(\hat{P}-m_l)\hat{e}^*\gamma^\nu\hat{p}_\pi
-\hat{D}_1(\hat{P}-m_l)\hat{e}^*(\hat{p}_l-m_l)\hat{p}_\pi \right.\\
\left.-2B_{2\mu\nu}\gamma^\mu\hat{e}^*\gamma^\nu+2\hat{B}_1\hat{e}^*\hat{P}+2\hat{p}_l\hat{e}^*\hat{B}_1
+8m_lB_{1\mu}e^{*\mu} \right.\\
\left.-2b_0(\hat{p}_l\hat{e}^*\hat{P}+4m_l(p_l
e^*)+m^2_l\hat{e}^*)\right]v_l;
\end{multline}

\begin{multline}
A_{5a-IR}=\frac{i}{8\pi^2}\sum_n
M_{5a}M_{IR}\rvert_{\substack{q\to k,\\
p_\pi\to k_\pi,\\ e\to e_k}}
=\frac{ie^2}{4\pi^2}\frac{G}{\sqrt{2}}\sin{\theta_c}ef_+(0)
\bar{u}_\nu\left\{(\hat{p}_K+\hat{P}')\right.\\
\left.\times \left[c'_0(p_l P')-a'_0\frac{p_\pi P'}{p_\pi
q}\right]-\left[\hat{C}'_1(p_l P')-a'_P\frac{p_\pi P'}{p_\pi
q}\hat{P}'\right]\right\}(1+\gamma_5)v_l\frac{p_\pi e^*}{p_\pi q};
\end{multline}

\begin{multline}
A_{5a-mag}=\frac{i}{8\pi^2}\sum_n
M_{5a}M_{mag}\rvert_{\substack{q\to k,\\
p_\pi\to k_\pi,\\ e\to e_k}}
=\frac{ie^2}{4\pi^2}\frac{G}{\sqrt{2}}\sin{\theta_c}ef_+(0)\\
\times\bar{u}_\nu(\hat{p}_K+\hat{P}')(1+\gamma_5)\hat{C}'_1\hat{P}' v_l \frac{p_\pi
e^*}{2p_\pi q};
\end{multline}

\begin{multline}
A_{5a-Low}=\frac{i}{8\pi^2}\sum_n
M_{5a}M_{Low}\rvert_{\substack{q\to k,\\
p_\pi\to k_\pi,\\ e\to e_k}}
=\frac{ie^2}{4\pi^2}\frac{G}{\sqrt{2}}\sin{\theta_c}ef_+(0)
\bar{u}_\nu\hat{P}'(1+\gamma_5)v_l \frac{p_\pi e^*}{p_\pi q}\\
\times \left[a'_0-a'_P\frac{p_\pi P'}{p_\pi q}\right];
\end{multline}

\begin{multline}
A_{5b-IR}=\frac{i}{8\pi^2}\sum_n
M_{5b}M_{IR}\rvert_{\substack{q\to k,\\
p_\pi\to k_\pi,\\ e\to e_k}}
=\frac{ie^2}{4\pi^2}\frac{G}{\sqrt{2}}\sin{\theta_c}ef_+(0)
\bar{u}_\nu\left\{(\hat{p}_K+\hat{P}')\right.\\
\times \left[\frac{p_\pi P'}{p_\pi q}(b'_0(p_\pi e^*)-B'_{1\mu}e^{*\mu})-\frac{p_\pi
e^*}{p_\pi q}(a'_0-a'_P)-(p_l
p_\pi)(d'_0(p_\pi e^*)-D'_{1\mu}e^{*\mu})\right]\\
-\left[\frac{p_\pi P'}{p_\pi q}(\hat{B}'_1(p_\pi e^*)-B'_{2\mu\nu}\gamma^\mu
e^{*\nu})-\frac{1}{p_\pi q}(a'_P(p_\pi e^*)\hat{P}'-A'_{2\mu\nu}\gamma^\mu
e^{*\nu})\right.\\
\left.\left.-(p_l p_\pi)((p_\pi e^*)\hat{D}'_1-D'_{2\mu\nu}\gamma^\mu
e^{*\nu})\right]\right\}(1+\gamma_5)v_l;
\end{multline}

\begin{multline}
A_{5b-mag}=\frac{i}{8\pi^2}\sum_n
M_{5b}M_{mag}\rvert_{\substack{q\to k,\\
p_\pi\to k_\pi,\\ e\to e_k}}
=\frac{ie^2}{4\pi^2}\frac{G}{\sqrt{2}}\sin{\theta_c}ef_+(0)
\bar{u}_\nu(\hat{p}_K+\hat{P}')(1+\gamma_5)\\
\times \frac{1}{2}\left[D'_{2\mu\nu}\gamma^\mu e^{*\nu}-(p_\pi
e^*)\hat{D}'_1\right]\hat{p}_\pi v_l;
\end{multline}

\begin{multline}
A_{5b-Low}=\frac{i}{8\pi^2}\sum_n
M_{5b}M_{Low}\rvert_{\substack{q\to k,\\
p_\pi\to k_\pi,\\ e\to e_k}}
=\frac{ie^2}{4\pi^2}\frac{G}{\sqrt{2}}\sin{\theta_c}ef_+(0)\\
\bar{u}_\nu\left[\frac{p_\pi P'}{p_\pi q}(\hat{B}'_1(p_\pi
e^*)-B'_{2\mu\nu}\gamma^\mu e^{*\nu})-\frac{1}{p_\pi q}(a'_P(p_\pi
e^*)\hat{P}'-A'_{2\mu\nu}\gamma^\mu
e^{*\nu})\right.\\
\left.-\hat{p}_\pi((p_\pi e^*)b'_0-B'_{1\mu}e^{*\mu})\right](1+\gamma_5)v_l;
\end{multline}

\begin{multline}
A_{5c-IR}=\frac{i}{8\pi^2}\sum_n
M_{5c}M_{IR}\rvert_{\substack{q\to k,\\
p_\pi\to k_\pi,\\ e\to e_k}}
=\frac{ie^2}{4\pi^2}\frac{G}{\sqrt{2}}\sin{\theta_c}ef_+(0)
\bar{u}_\nu\left\{(\hat{p}_K+\hat{P}')\right.\\
\left.\times \left[\frac{p_\pi e^*}{p_\pi q}(a'_0-a'_P)-(p_l
e^*)c'_0\right]-\left[\frac{p_\pi e^*}{p_\pi q}a'_P\hat{P}'-\frac{1}{p_\pi
q}A'_{2\mu\nu}\gamma^\mu e^{*\nu}-(p_l e^*)\hat{C}'_1\right]\right\}(1+\gamma_5)v_l;
\end{multline}

\begin{multline}
A_{5c-mag}=\frac{i}{8\pi^2}\sum_n
M_{5c}M_{mag}\rvert_{\substack{q\to k,\\
p_\pi\to k_\pi,\\ e\to e_k}}
=\frac{ie^2}{4\pi^2}\frac{G}{\sqrt{2}}\sin{\theta_c}ef_+(0)
\left[-\frac{1}{2}\bar{u}_\nu(\hat{p}_K+\hat{P}')(1+\gamma_5)\hat{C}'_1\hat{e}^*
v_l\right];
\end{multline}

\begin{multline}
A_{5c-Low}=\frac{i}{8\pi^2}\sum_n
M_{5c}M_{Low}\rvert_{\substack{q\to k,\\
p_\pi\to k_\pi,\\ e\to e_k}}
=\frac{ie^2}{4\pi^2}\frac{G}{\sqrt{2}}\sin{\theta_c}ef_+(0)\\
\times \bar{u}_\nu\left[\frac{p_\pi e^*}{p_\pi q}a'_P\hat{P}'-\frac{1}{p_\pi
q}A'_{2\mu\nu}\gamma^\mu e^{*\nu}-a'_0\hat{e}^*\right](1+\gamma_5)v_l;
\end{multline}

\begin{multline}
A_{7-IR}=\frac{i}{8\pi^2}\sum_n
M_{7}M_{IR}\rvert_{\substack{p_l\to k_l,\\ p_\pi \to k_\pi}}
=\frac{ie^2}{8\pi^2}\frac{G}{\sqrt{2}}\sin{\theta_c}ef_+(0)\bar{u}_\nu(1-\gamma_5)
(\hat{p}_K+\hat{P}''+m_l)\\
\times \left[\left(E''_{2\mu\nu}+F''_{2\mu\nu}\right)\gamma^\mu e^{*\nu}-m_l
E''_{1\mu}e^{*\mu}-(\hat{P}''-m_l)F''_{1\mu}e^{*\mu}\right]
(\hat{p}_\pi+\hat{P}''+m_l) v_l;
\end{multline}

\begin{multline}
A_{7-mag}=\frac{i}{8\pi^2}\sum_n
M_{7}M_{mag}\rvert_{\substack{p_l\to k_l,\\ p_\pi \to k_\pi}}
=\frac{ie^2}{8\pi^2}\frac{G}{\sqrt{2}}\sin{\theta_c}ef_+(0)\frac{1}{2}\bar{u}_\nu(1-\gamma_5)\\
\times \left[(\hat{p}_K+\hat{P}'')\hat{q}\hat{e}^* \left(\hat{E}''_1-m_l
e''_0\right)
+m_l\hat{E}''_1\hat{q}\hat{e}^*-E''_{2\mu\nu}\gamma^\mu\hat{q}\hat{e}^*\gamma^\nu\right]
(\hat{p}_\pi+\hat{P}''+m_l) v_l;
\end{multline}

\begin{multline}
A_{7-Low}=\frac{i}{8\pi^2}\sum_n
M_{7}M_{Low}\rvert_{\substack{p_l\to k_l,\\ p_\pi \to k_\pi}}
=\frac{ie^2}{8\pi^2}\frac{G}{\sqrt{2}}\sin{\theta_c}ef_+(0)\bar{u}_\nu(1-\gamma_5)\\
\times \left[F''_{2\mu\nu}\hat{q}\gamma^\mu
e^{*\nu}-\hat{q}(\hat{P}''-m_l)F''_{1\mu}e^{*\mu}+\hat{e}^*\hat{D}''_1\right]
(\hat{p}_\pi+\hat{P}''+m_l) v_l;
\end{multline}

\begin{multline}
A_{9a-10}=\frac{i}{8\pi^2}\sum_n M_{9a}M_{10}
=\frac{ie^3}{8\pi^2}\frac{G}{\sqrt{2}}\sin{\theta_c}f_+(0)\bar{u}_\nu(1-\gamma_5)
(\hat{p}_K+\hat{P}'''+m_l)\\
\times \left(\hat{D}'''_1-m_ld'''_0\right)(\hat{p}_\pi+\hat{P}'''+m_l)
\left(\frac{p_l e^*}{p_l q}+\frac{\hat{q}\hat{e}^*}{2p_l q}\right) v_l;
\end{multline}

\begin{multline}
A_{9b-10}=\frac{i}{8\pi^2}\sum_n M_{9b}M_{10}
=-\frac{ie^3}{8\pi^2}\frac{G}{\sqrt{2}}\sin{\theta_c}f_+(0)\bar{u}_\nu(1-\gamma_5)
(\hat{p}_K+\hat{P}'''+m_l)\\
\times \left[\left(F'''_{2\mu\nu}\gamma^\mu e^{*\nu}-m_l F'''_{1\mu}e^{*\mu}\right)
(\hat{p}_\pi+\hat{P}'''+m_l)\right. +\frac{1}{2}\left(\hat{F}'''_1-m_lf'''_0\right)
\hat{q}\hat{e}^*(\hat{p}_\pi+\hat{P}''')\\
\left.+\frac{1}{2}m_l\hat{q}\hat{e}^*\hat{F}'''_1
-\frac{1}{2}F'''_{2\mu\nu}\gamma^\mu\hat{q}\hat{e}^*\gamma^\nu\right] v_l;
\end{multline}

\begin{multline}
A_{9c-10}=\frac{i}{8\pi^2}\sum_n M_{9c}M_{10}
=-\frac{ie^3}{8\pi^2}\frac{G}{\sqrt{2}}\sin{\theta_c}f_+(0)\bar{u}_\nu(1-\gamma_5)
(\hat{p}_K+\hat{P}'''+m_l)\\
\times
\left((\hat{P}'''-m_l)e'''_0-\hat{E}'''_1\right)(\hat{p}_\pi+\hat{q}+\hat{P}'''+m_l)v_l
\frac{p_\pi e^*}{p_\pi q};
\end{multline}

\begin{multline}
A_{9d-10}=\frac{i}{8\pi^2}\sum_n M_{9d}M_{10}
=\frac{ie^3}{8\pi^2}\frac{G}{\sqrt{2}}\sin{\theta_c}f_+(0)\bar{u}_\nu(1-\gamma_5)
(\hat{p}_K+\hat{P}'''+m_l)\\
\times \left[(\hat{P}'''-m_l)G'''_{1\mu}e^{*\mu}-G'''_{2\mu\nu}\gamma^\mu
e^{*\nu}\right](\hat{p}_\pi-\hat{q}+\hat{P}'''+m_l)v_l;
\end{multline}

\begin{multline}
A_{9e-10}=\frac{i}{8\pi^2}\sum_n M_{9e}M_{10}
=\frac{ie^3}{8\pi^2}\frac{G}{\sqrt{2}}\sin{\theta_c}f_+(0)\cdot
2\bar{u}_\nu(1-\gamma_5)(\hat{p}_K+\hat{P}'''+m_l)\\
\times \left((\hat{P}'''-m_l)e'''_0-\hat{E}'''_1\right)\hat{e}^*v_l.
\end{multline}

\end{document}